\documentclass[useAMS,usenatbib]{mn2e}

\usepackage{graphicx}

\usepackage{amsmath}
\usepackage{amssymb}

\newcommand{\apj}{ApJ}

\newcommand{\aj}{AJ}
\newcommand{\aap}{A\&A}
\newcommand{\aaps}{A\&AS}
\newcommand{\apss}{Ap\&SS}
\newcommand{\mnras}{MNRAS}

\newcommand{\ion}[2]{{#1}\,{\sc {\small{#2}}}}          

\title[Combined analysis of HD\,183648]{HD\,183648: a \textbf {\textit{Kepler}} eclipsing binary with anomalous ellipsoidal variations and a pulsating component}
\author[T. Borkovits et al.]{T. Borkovits$^{1,2,3}$\thanks{E-mail:
borko@electra.bajaobs.hu (TB)}, A. Derekas$^{2,3}$, J. Fuller$^4$, Gy. M. Szab\'o$^{2,3}$, K. Pavlovski$^5$,
\newauthor  B. Cs\'ak$^3$, \'A. D\'ozsa$^3$, J. Kov\'acs$^3$, R. Szab\'o$^{2}$, K. M. Hambleton$^{6,7}$, 
\newauthor  K. Kinemuchi$^{8}$, V. Kolbas$^5$, D. W. Kurtz$^{6}$, F. Maloney$^{7}$, 
\newauthor A. Pr\v{s}a$^{7}$, J. Southworth$^{9}$, J. Sztakovics$^{10}$, I. B. B\'\i r\'o$^{1}$, I. Jankovics$^3$\\
$^{1}$Baja Astronomical Observatory, H-6500 Baja, Szegedi \'ut, Kt. 766, Hungary\\
$^{2}$Konkoly Observatory, MTA CSFK, H-1121 Budapest, Konkoly Thege M. \'ut 15-17, Hungary\\
$^{3}$ELTE Gothard Astrophysical Observatory, H-9704 Szombathely, Szent Imre herceg \'ut 112, Hungary\\
$^{4}$Department of Astronomy, Center for Space Research, Cornell University, Ithaca, NY 14853, USA\\
$^{5}$Department of Physics, University of Zagreb, Bijeni\v cka cesta 32, 10000 Zagreb, Croatia\\
$^{6}$Jeremiah Horrocks Institute, University of Central Lancashire, Preston PR1\,2HE, UK\\
$^{7}$Department of Astrophysics and Planetary Science, Villanova University, 800 Lancaster Ave, Villanova PA 19085, USA\\
$^{8}$Apache Point Observatory, Sunspot NM 88349, USA\\
$^{9}$Astrophysics Group, Keele University, Newcastle-under-Lyme, ST5 5BG, UK \\
$^{10}$Astronomical Department of E\"otv\"os University, H-1118 P\'azm\'any P\'eter stny. 1/A, Budapest, Hungary}

\begin{document}

\date{Accepted ??? Received ???; in original form ???}


\maketitle

\label{firstpage}

\begin{abstract}
KIC\,8560861 (HD\,183648) is a marginally eccentric ($e=0.05$) eclipsing binary with an orbital period of $P_\mathrm{orb}=31.973$\,d, exhibiting mmag amplitude pulsations on time scales of a few days. We present the results of the complex analysis of high and medium-resolution spectroscopic data and {\it Kepler} Q0 -- Q16 long cadence photometry. The iterative combination of spectral disentangling, atmospheric analysis, radial velocity and eclipse timing variation studies, separation of pulsational features of the light curve, and binary light curve analysis led to the accurate determination of the fundamental stellar parameters. We found that the binary is composed of two main sequence stars with an age of $0.9\pm0.2$\,Gyr, having masses, radii and temperatures of $M_1=1.93\pm0.12$\,M$_{\odot}$, $R_1=3.30\pm0.07$\,R$_{\odot}$, $T_\mathrm{eff1}=7650\pm100$\,K for the primary, and $M_2=1.06\pm0.08$\,M$_{\odot}$, $R_2=1.11\pm0.03$\,R$_{\odot}$, $T_\mathrm{eff2}=6450\pm100$\,K for the secondary. After subtracting the binary model, we found three independent frequencies, two of which are separated by twice the orbital frequency. We also found an enigmatic half orbital period sinusoidal variation that we attribute to an anomalous ellipsoidal effect. Both of these observations indicate that tidal effects are strongly influencing the luminosity variations of HD\,183648. The analysis of the eclipse timing variations revealed both a parabolic trend, and apsidal motion with a period of $P_\mathrm{apse}^\mathrm{obs}=10\,400\pm3\,000$\,y, which is three times faster than what is theoretically expected. These findings might indicate the presence of a distant, unseen companion.
\end{abstract}


\section{Introduction}

Eclipsing binary stars have long been recognized as key objects for calibrating astronomical observations in terms of fundamental stellar parameters. In fact, binarity has been, until recently, the only way to accurately determine stellar masses. The combination of time-series photometry and spectroscopy of eclipsing binaries enables us to measure the most accurate masses and radii for stars, namely to better than 1\,per\,cent \citep[e.g.][]{and91,cla08,tor10}.

There is a special group of eclipsing binaries that take a very important place in astrophysics: those with pulsating components. Such systems are important laboratories for confronting theories with observations. The mass measured from an eclipsing binary can be compared with those coming from other determinations and models, such as evolutionary or pulsational models \citep{aer07}. 

Eclipses can be helpful in mode detection and identification, and pulsations enable us to measure the internal rotational velocity of the pulsating star through the rotational splitting of the non-radial modes \citep{bap93,gou00,gam03,mik05,bir11}. In close binary systems it is common that tidal forces induce pulsations \citep{wil03,wel11,tho12,ful13} and in special cases resonances of the frequency of the pulsation and the orbital period can be detected \citep[][Hambleton~2014, in preparation]{ful13,ham13}.

Almost every type of pulsating star has been found as a component in an eclipsing binary system. A good overview of these systems and their distribution is given by \citet{pig06}. Since then the number of known systems has grown significantly thanks to large ground base photometric surveys \citep[e.g.][]{pig07,mic08} and the unprecedented quality of the photometric light curves delivered by space telescopes CoRoT \citep{mac13,dasilvaetal14} and {\it Kepler} \citep{ost10,der11,sou11,tel12,debosscheretal13,fra13,ham13,bec14,maceronietal14}.

Here we present the analysis of an eclipsing binary with a pulsating component discovered in the {\it Kepler} dataset. KIC\,8560861 (HD\,183648) is a relatively long-period ($P_\mathrm{orb} = 31.97$\,d), marginally eccentric ($e = 0.05$) eclipsing binary system which exhibits mmag pulsations with periods on the order of a few days. It has a magnitude of $V = 8.5$, hence it is above the saturation limit of the {\it Kepler} space telescope, which was taken into account for the data reduction (see in Sect.\ \ref{Sect2}). It is listed in the catalogues of the first and second releases of the Kepler Eclipsing Binary Catalogue \citep{prs11,sla11}. The Kepler Input Catalogue (KIC) lists the following parameters for this target: $r_\mathrm{SDSS}=8.498$, $T_\mathrm{eff} = 7647$\,K, $\log{g} = 3.532$, $\mathrm{[Fe/H]}=-0.084$.

In the following sections we present the combined analysis of the {\it Kepler} photometry and ground based spectroscopic data, which includes (i) analysis of eclipse timing variation (Sect.\,\ref{Sect:ETV}), (ii) determination of atmospheric properties of the primary star from crosscorrelation function spectroscopy (Sect.\,\ref{Subsect.:fundparam}), (iii) a radial velocity study (Sect.\,\ref{Subsect:radvelstudy}), $(iv)$ spectral disentangling and determination of the dynamical masses (Sect.\,\ref{sec: spd}), $(v)$ eclipse light curve analysis (Sect.\ref{lcanalysis}), and $(vi)$ determination of the pulsation frequencies (Sect.\,\ref{Subsect:freqsearch}). Finally, in Sect.\,\ref{Subsect:tidaloscillation} we discuss the characteristics and the possible tidal origin of the detected oscillations.

\section[]{Observations and data reduction}
\label{Sect2}

\subsection{{\it Kepler} photometry}

The photometric analysis is based on photometry from the {\it Kepler} space telescope \citep{bor10,gil10,koc10,jen10a,jen10b}. HD\,183648 was observed both in long and short cadence mode between 2009 and 2013. The long cadence (time resolution of 29.4\,min) dataset covers nearly the whole length of Kepler's 4-y life-time (Quarters 0 -- 16), while it was observed for only 30\,d in Q3.2 in short cadence (time resolution of 58.9\,s) mode. 

The MAST\footnote{http://archive.stsci.edu/kepler/} database indicates $0.1-0.3$\,per\,cent contamination, depending on the quarter in question. We have downloaded the target pixel files for all quarters and performed several checks by using PyKE\footnote{http://keplerscience.arc.nasa.gov/PyKE.shtml} tools. First, we verified that all signals come from one object; that is, no contamination is seen from a close-by blended object within Kepler's resolution (4\,arcsec). This was done by examining the amplitude of the (visible) signals in individual pixels. Any signal coming from a different source would be revealed by a displaced pixel showing a higher amplitude of that signal. We also checked that no signal was lost due to the assigned target aperture mask.  Due to the brightness of the star and saturation on the {\it Kepler} photometer, the target aperture mask was elongated. Along the elongation axis, we still detect signal from the star, and we assume the flux could be still detected in pixels outside the target aperture. However, we have determined that the contribution of these peripheral pixels outside of the target aperture is negligible. We estimate that the lost flux from the star is less than 0.01\,per\,cent.

\subsection{Spectroscopy}
\label{Subsect:Spectroscopy}

We obtained high and medium resolution spectra at five observatories. We took 2 spectra in 2010 at Kitt Peak National Observatory (KPNO), USA, using the Echelle Spectrograph at the Mayall 4-meter telescope with a resolution of $R = 20\,000$ in the spectral range $4700 - 9300$\,\AA. 34 spectra were taken on 11 nights in 2012 with the eShel spectrograph mounted on a 0.5-m Ritchey-Chr\'etien telescope at the Gothard Astronomical Observatory (GAO), Szombathely, Hungary, in the spectral range $4200 - 8700$\,\AA\ with a resolution of $R = 11\,000$. The wavelength calibration was done using a ThAr lamp. The same instrument was used at Piszk\'estet\H o Observatory (PO), Hungary mounted on the 1-m telescope, where we took 36 spectra on 16 nights in 2012 and 2013. A detailed description of the instrument can be found in \citet{csa14}. We obtained 5 spectra at Apache Point Observatory (APO), USA, using the ARCES Echelle spectrograph on the 3.5~m telescope with a resolution of $R = 31\,500$ in the spectral range $3200 - 10\,000$\,\AA. We took 3 spectra at Lick Observatory, USA, using the Hamilton Echelle Spectrograph mounted on the Shane 3-meter Telescope. The resolution was $R = 37\,000$ in the spectral range  $4200 - 6850$\,\AA. The journal of observations can be found in Table\,\ref{obsjour}.

\begin{table}  
\begin{center}
\caption{Journal of observations.} 
\label{obsjour}  
\begin{tabular}{|llcc|} 
\hline 
Observatory &  wavelength range & Res.  & No. of spectra \\
\hline
GAO & 4200--8700 \AA & 11 000 & 34 \\
Piszk\'estet\H o & 4200--8700 \AA & 11 000 & 36 \\
KPNO &  4700--9300 \AA & 20 000 & 2 \\
APO & 3200--10000 \AA & 31 500 & 5 \\
LICK &  4200--6850 \AA & 37 000 & 3  \\
\hline   
\end{tabular} 
\end{center}  
\end{table}

All spectra were reduced either using IRAF or a dedicated pipeline, then normalised to the continuum level. The radial velocities were determined by cross-correlating the spectra with a well-matched theoretical template spectrum from the extensive spectral library of \citet{mun05}. In cases of spectra obtained at Gothard Astronomical Observatory and Piszk\'estet\H o Observatory, we co-added those taken in the same night to produce higher signal-to-noise ratios. All radial velocities were corrected to barycentric radial velocities, and are listed in Table\,\ref{Tab:radveldata}. 

By the use of this conventional cross-correlation technique, HD\,183648 was found to be a single lined spectroscopic binary (SB1), which was in good agreement with the expectation based on the preliminary light curve fit.

\begin{table}
\caption{Radial velocity measurements }
\label{Tab:radveldata}
\begin{tabular}{@{}lr|lr}
\hline
  BJD & \multicolumn{1}{c}{$v_\mathrm{rad}$}  & BJD & \multicolumn{1}{c}{$v_\mathrm{rad}$} \\
  $-2\,400\,000$ & (km\,s$^{-1}$) &   $-2\,400\,000$ & (km\,s$^{-1}$) \\
\cline{1-4}
\multicolumn{2}{c}{GAO}     &  $56057.48133$   & $-23.0(5)$ \\ \cline{1-2}
$55987.67491^*$ & $10.5(4)$  & $56058.45077$   & $-25.8(5)$ \\
$56009.64102$   & $11.3(3)$  & $56059.49054$   & $-30.1(5)$   \\
$56012.60284$   & $25.5(5)$  & $56514.36738$   & $-20.2(3)$ \\
$56015.54585$   & $27.6(5)$  & $56516.55074$   & $-9.7(4)$  \\
$56020.52020$   & $5.0(4)$   & $56521.53346^*$ & $19.0(3)$  \\
$56084.43023$   & $6.4(5)$   & $56542.35874$ & $-30.0(3)$ \\
$56091.37875$   & $-32.0(4)$ & $56555.37107$ & $24.9(3)$  \\ \cline{3-4}
$56104.40849$   & $6.5(5)$   &  \multicolumn{2}{c}{KPNO}\\ \cline{3-4}
$56105.52976$   & $15.7(5)$  & $55738.95801^*$ & $-26.0(2)$\\
$56106.49568$   & $20.4(5)$  & $55742.97074^*$ & $-32.4(2)$\\ \cline{3-4}
$56117.45252$   & $0.3(5)$   &  \multicolumn{2}{c}{APO} \\ \cline{1-2} \cline{3-4}
\multicolumn{2}{c}{PO} & $56126.85215^{*}$ & $-29.2(2)$ \\ \cline{1-2}
$55990.65432^*$ & $-3.6(4)$ &  $56204.66367^*$ & $29.2(2)$ \\ 
$55995.67350^*$ & $-29.9(3)$ &  $56224.65775^*$ & $-26.2(2)$ \\
$55996.65663$   & $-28.2(4)$ & $56225.65443^*$ & $-22.9(2)$ \\ 
$55998.65543^*$ & $-34.6(3)$ &  $56228.58654^*$ & $-9.9(2)$  \\ \cline{3-4}
$56000.65120$   & $-29.0(5)$  & \multicolumn{2}{c}{Lick} \\  \cline{3-4}
$56008.64047$   & $9.3(3)$  & $56133.00141^*$ & $-11.0(2)$ \\
$56048.52003^*$ & $25.7(3)$  & $56133.99831^*$ & $-4.0(2)$\\
$56053.49436$   & $0.9(5)$  & $56134.87329^*$ & $3.3(2)$  \\
\hline
 \end{tabular}

$^*$: measurements used for spectral disentanglement \\
\end{table}

\begin{table*}
\caption{Times of minima of HD\,183648, after removal of the oscillations from the light curve (see text for details). Half-integer cycle numbers refer to secondary minima.}
 \label{Tab:ToM}
 \begin{tabular}{@{}lrlllllll}

  \hline
  BJD & Cycle  & std. dev. & BJD & Cycle  & std. dev. & BJD & Cycle  & std. dev. \\
  $-2\,400\,000$ & no. &   \multicolumn{1}{c}{$(d)$} & $-2\,400\,000$ & no. &   \multicolumn{1}{c}{$(d)$} & $-2\,400\,000$ & no. &   \multicolumn{1}{c}{$(d)$} \\
\hline
54966.868878 &  0.0 & 0.000085 & 55463.197910 & 15.5 & 0.000073 &  55942.797650 & 30.5 &  0.000179 \\
54983.600442 &  0.5 & 0.000193 & 55478.440878 & 16.0 & 0.000064 &  55958.039172 & 31.0 &  0.000085 \\
55030.815327 &  2.0 & 0.000086 & 55495.171972 & 16.5 & 0.000070 &  55974.770697 & 31.5 &  0.000192 \\
55047.547225 &  2.5 & 0.000184 & 55510.413023 & 17.0 & 0.000050 &  55990.013154 & 32.0 &  0.000084 \\
55062.788056 &  3.0 & 0.000083 & 55527.144610 & 17.5 & 0.000095 &  56006.743789 & 32.5 &  0.000194 \\
55079.519265 &  3.5 & 0.000197 & 55542.386918 & 18.0 & 0.000058 &  56021.986202 & 33.0 &  0.000084 \\
55094.761357 &  4.0 & 0.000084 & 55574.359507 & 19.0 & 0.000070 &  56038.716950 & 33.5 &  0.000180 \\
55111.494290 &  4.5 & 0.000193 & 55591.091868 & 19.5 & 0.000059 &  56053.959844 & 34.0 &  0.000086 \\
55126.734729 &  5.0 & 0.000084 & 55606.333366 & 20.0 & 0.000084 &  56070.690183 & 34.5 &  0.000195 \\
55143.466542 &  5.5 & 0.000191 & 55623.065473 & 20.5 & 0.000191 &  56085.932769 & 35.0 &  0.000085 \\
55158.708288 &  6.0 & 0.000084 & 55655.037652 & 21.5 & 0.000193 &  56102.664102 & 35.5 &  0.000194 \\
55175.440176 &  6.5 & 0.000200 & 55670.279778 & 22.0 & 0.000085 &  56117.906765 & 36.0 &  0.000090 \\
55190.681240 &  7.0 & 0.000079 & 55687.011344 & 22.5 & 0.000187 &  56134.637722 & 36.5 &  0.000193 \\
55207.413066 &  7.5 & 0.000178 & 55702.252701 & 23.0 & 0.000084 &  56149.879518 & 37.0 &  0.000084 \\
55222.654283 &  8.0 & 0.000084 & 55718.984941 & 23.5 & 0.000195 &  56166.611156 & 37.5 &  0.000181 \\
55239.385793 &  8.5 & 0.000193 & 55734.226527 & 24.0 & 0.000085 &  56181.853449 & 38.0 &  0.000086 \\
55254.627387 &  9.0 & 0.000084 & 55750.957816 & 24.5 & 0.000206 &  56198.583760 & 38.5 &  0.000195 \\
55271.359097 &  9.5 & 0.000192 & 55766.199574 & 25.0 & 0.000086 &  56213.826767 & 39.0 &  0.000085 \\
55286.600691 & 10.0 & 0.000084 & 55782.931099 & 25.5 & 0.000182 &  56230.558449 & 39.5 &  0.000193 \\
55303.332429 & 10.5 & 0.000101 & 55798.173244 & 26.0 & 0.000081 &  56245.800354 & 40.0 &  0.000730 \\
55318.573707 & 11.0 & 0.000081 & 55814.904382 & 26.5 & 0.000196 &  56262.530887 & 40.5 &  0.000194 \\
55335.305117 & 11.5 & 0.000051 & 55830.146219 & 27.0 & 0.000086 &  56277.773536 & 41.0 &  0.000085 \\
55350.547360 & 12.0 & 0.000043 & 55846.877296 & 27.5 & 0.000193 &  56294.504414 & 41.5 &  0.000180 \\
55367.279166 & 12.5 & 0.000076 & 55862.120139 & 28.0 & 0.000084 &  56309.747978 & 42.0 &  0.000093 \\
55382.520503 & 13.0 & 0.000092 & 55878.850805 & 28.5 & 0.000193 &  56326.478929 & 42.5 &  0.000182 \\
55399.251293 & 13.5 & 0.000106 & 55894.092879 & 29.0 & 0.000084 &  56341.720533 & 43.0 &  0.000085 \\
55414.494091 & 14.0 & 0.000057 & 55910.823884 & 29.5 & 0.000194 &  56373.694490 & 44.0 &  0.000085 \\
55431.225013 & 14.5 & 0.000029 & 55926.067023 & 30.0 & 0.000085 &  56390.425555 & 44.5 &  0.000194 \\
55446.466863 & 15.0 & 0.000086 &              &      &          &  		&      &	   \\
\hline
 \end{tabular}

\end{table*}

\section[]{Eclipse timing analysis}
\label{Sect:ETV}

In the case of an eclipsing binary, eclipse timing analysis is the most powerful tool for $(i)$ determining an accurate period, $(ii)$ detecting and identifying any kind of period variation, either physical or apparent, (iii) calculating an accurate value of the $e\cos\omega$ parameter for eccentric systems, and ($iv$) detecting a slow variation in the eclipse times caused by an apsidal advance of the binary's orbit. 

We therefore analysed the eclipse timing variations (ETV) first. The individual times of minima were determined with the following algorithm. First a folded, equally binned and averaged light curve was formed from the whole Q0 -- Q16 data set. Then two template minima were calculated with polynomial fits of degree $4-6$ on the primary and secondary eclipses. Finally, these templates were fitted to all individual minima. As an alternative method and check, parabolic and cubic linear least-squares fits and minima determinations to each individual minimum were also applied. These methods are very similar to those used by \citet{rappaportetal13}, for example. Furthermore, we estimated the accuracy of minima determinations by calculating the standard deviations for each minimum with bootstrap sampling \citep[see, e.g.,][]{bratetal14}.  

Our observed times of minima ($O$) were compared with calculated times of minima ($C$)  with the following linear ephemeris to give values of $O-C$:
\begin{equation}
\mathrm{MIN}_\mathrm{I}=2454966.8687+31.9732\times E,
\label{Eq:linephem}
\end{equation}
which was determined with the method described above.

\begin{figure}
\includegraphics[width=84mm]{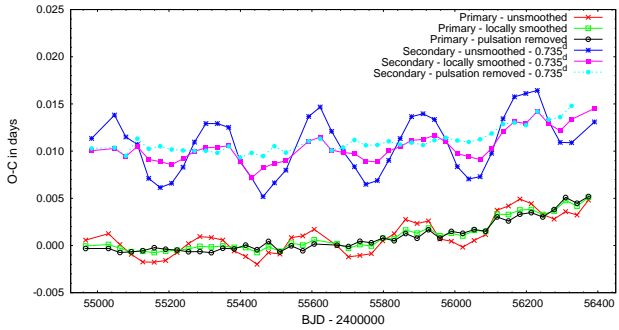}
\caption{$O-C$ diagram of the Eclipse Timing Variations (ETV) for the primary and secondary minima. (For better visibility the secondary curve is shifted by $0.735$ days, which corresponds to the displacement of secondary minima from phase 0.5.) The parabolic trend seems to be real. The cyclic feature, however, arises from the pulsational distortion of the light curve. The apparent cyclic variation can be eliminated with either a local smoothing of the light curve around each minimum, or the removal of the pulsational variations. (See text for details.)}
\label{Fig:ETV}
\end{figure}

The raw $O-C$ diagram revealed a complex nature which was a combination of a cyclic variation with a period of $\sim287$\,d and a slower, quadratic term (red and blue curves in Fig.\,\ref{Fig:ETV}). The primary and the secondary minima correlated in both features; however, the cyclic variation had a greater amplitude in the secondary curve. In the present situation this variation does not arise from the presence of a third companion (which is the most common interpretation of such $O-C$ curves), nor does it arise from any other real physical or geometric cause. It is purely the result of the pulsational distortion of the light curve. This comes from the fact that the mean pulsational frequency is close to a $165:9$ ratio to the orbital frequency (see Sect.\,\ref{lcanalysis}), and consequently, every ninth primary, and secondary eclipsing minima are distorted in a similar way. 

Apparent ETVs forced by light curve variations are also seen in other stars using accurate {\it Kepler} data. Recently, \citet{tranetal13} reported quasi-periodic $O-C$ diagrams for almost 400 short period, mostly overcontact binaries, whose phenomena were interpreted as an effect of large, spotted regions on the binary members. A difference, however, is that the spotted stars resulted in anticorrelated primary and secondary ETVs, while in the present case the ETVs are correlated.

This apparent timing variation was eliminated by the removal of the pulsations from the light curve. In Fig.\,\ref{Fig:ETV} we illustrate this in two different ways. First we applied local smoothing around all individual minima. We fitted polynomials of order $4-8$ on short sections of the light curve before the first and after the fourth contacts, and then removed them from the light curves (as was done by \citealt{borkovitsetal13}). This procedure removed the effect within the errors from the primary $O-C$ curve, but some residuals with moderate amplitude remained in the secondary one. Next, after the light curve analysis, we removed the pulsations from the original light curve in the manner described in Sect.\,\ref{lcanalysis}. Running our code on the latter, prewhitened curve, we obtained both primary and secondary $O-C$ curves without the cyclic variations. 

\begin{figure}
\includegraphics[width=84mm]{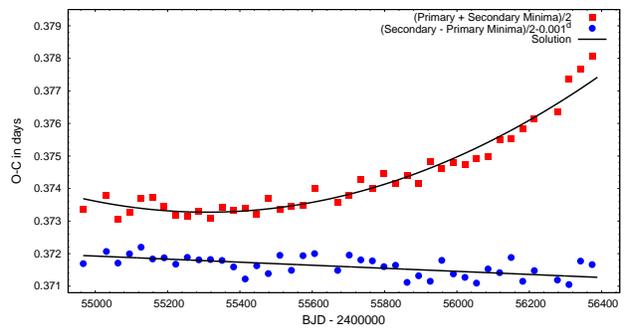}
\caption{The sum (red) and the difference (blue) of the primary and secondary $O-C$ curves calculated from data after the pulsations were removed, together with Levenberg-Marquardt fits (black lines). Such a visualization helps to separate the parabolic trend having correlated nature between primary and secondary minima, and apsidal motion which has an approximately anticorrelated effect for primary and secondary minima variations. The parabolic trend in the `sum' curve is evident. The small non-horizontal slope of the `difference' curve is an indication of apsidal motion.}
\label{Fig:ETVsumdifffit}
\end{figure}

For the final ETV analysis these latter, pulsation-removed $O-C$ curves were used. These prewhitened curves showed an additional feature. Subtracting the secondary $O-C$ values from the primary ones (which technically was carried out by a cubic spline interpolation of the secondary minima data to the times of the primary minima), we found that the difference curve had a non-zero slope, as can be seen in Fig.\,\ref{Fig:ETVsumdifffit}. This is a likely indication of apsidal motion in the binary. 

Therefore, we modelled the ETV in the following mathematical form:
\begin{eqnarray}
 \Delta&=&T(E)-T(0)-P_\mathrm{s}E \nonumber \\
&=&\sum_{i=0}^2c_iE^i+\frac{P_\mathrm{a}}{2\pi}\left[2\arctan\left(\frac{\mp e\cos\omega}{1+\sqrt{1-e^2}\pm e\sin\omega}\right)\right. \nonumber \\
&&\left.\mp\sqrt{1-e^2}\frac{e\cos\omega}{1\pm e\sin\omega}\right]_0^E,
\end{eqnarray}

where
\begin{equation}
\omega(E)=\omega(0)+\Delta\omega E.
\end{equation}

Here, in the first row, $T(E)$ means the observed time of the $E$-th minimum, $T(0)=T_0$ is the same for the reference minimum,  $P_\mathrm{s}$ stands for the sidereal (eclipsing) period. Note, the cycle number $E$ takes integer values for primary, and half-integer ones for secondary minima, respectively. In the second and third rows, the $c_0$, $c_1$ coefficients of the quadratic polynomial give corrections in $T_0$, $P_\mathrm{s}$, respectively, while $c_2$ is equal to the half of the constant period variation rate per cycle (i.e. $\Delta P/2$). The last two terms give the apsidal motion contribution. Usually it is given in the form of trigonometric series of $\omega$ \citep[see e.g.][]{gimenezgarcia83}. The present computational facilities, however, allow us to use its exact, analytic form. In this expression $P_\mathrm{a}\sim P_\mathrm{s}(1+\Delta\omega/2\pi)$ denotes the anomalistic period, $e$ stands for the eccentricity, while $\omega$ refers to the argument fo periastron of the primary's physical (or spectroscopic) orbit. This latter quantity varies in time. $\omega(0)=\omega_0$ means its value at $T_0$ epoch, and $\Delta\omega$ denotes the apsidal advance rate for one binary revolution. Furthermore, upper signs refer to primary, and lower ones to secondary minima, respectively.
%
%
Note, we neglect the small effects of the weak inclination dependence on the time of the deepest eclipse in eccentric binaries \citep[see e.g.][]{gimenezgarcia83}, and the intrinsic light-time effect between primary and secondary minima for stars of unequal masses \citep{fabrycky10}\footnote{Eq.\,(14) of the cited paper is valid strictly for $i = 90^{\circ}$; otherwise, it should be multiplied by $\sin i$ for the correct value.}. 

In order to determine the parameters listed above, the $\Delta$ function was fitted by a Levenberg-Marquardt-based differential correction procedure. For such a short time-scale, however, the $\Delta\omega$ parameter is highly correlated with $e$ and $\omega$ \citep[See][for details.]{claret98} Consequently, we decided to fix one of these three parameters, and adjust only the other two (together with the three polynomial coefficients $c_i$-s) in the differential correction process. Therefore, we fixed the eccentricity on its RV analysis obtained value. Then, in order to estimate the uncertainty of the parameters obtained, we repeated the process with slightly modified eccentricities. This refinement allowed us to reduce the uncertainty in the eccentricity an order of magnitude. The results of the complete process are listed in Table\,\ref{Tab:ETVresult}. In Fig.\,\ref{Fig:ETVsumdifffit} we plot our results on the averaged (red) and the difference (blue) $O-C$ curves. The first was calculated by summing the $O-C$ values of primary and secondary minima, while the second by with subtracting them. (The results were divided by two in both cases.) The advantage of such visualization is that it nicely separates quadratic variations and apsidal motion, as the former has correlated nature, while the second one shows primarily anticorrelated behaviour with respect to primary and secondary minima.

\begin{table}
 \caption{Results of ETV solution (one sigma uncertainties in the last digits are given in parentheses)}
 \label{Tab:ETVresult}
 \begin{tabular}{@{}lc}
  \hline
  $T_0\mathrm{~(BJD)}$ & $2\,454\,966.86896(20)$  \\
  $P_\mathrm{s}\mathrm{~(days)}$ & $31.973126(18)$  \\
  $\Delta P\mathrm{~(days/cycle)}$ & $7.2(8)\times10^{-6}$ \\
  \hline
  $e$ & $0.0477(1)$  \\
  $\omega_0 (\degr)$ & $37.260(22)$ \\
  $P_\mathrm{apse}\mathrm{~(years)}$ & $10\,432(3\,033)$  \\
  \hline
 \end{tabular}
\end{table}

Parabolic-shaped ETV curves, corresponding to constant period variations in time (or, more strictly, in cycle number), have been observed in hundreds of eclipsing binaries (see \citealp{sterken05}, in general, and \citealp{zhuetal12}, for a recent example). However, the most common interpretations, such as mass exchange, mass loss and magnetic interactions, can be neglected in this widely separated and therefore weakly interacting binary. Thus, in our case, the most probable source of the observed small period increase would be a gravitationally bound, distant, third companion. This additional component must be a faint object, as there is no evidence for an additional light source in the spectroscopic or the photometric data (see subsequent sections). There might be, however, weak indirect evidence for the presence of this body in the observed period of $P_\mathrm{apse}^\mathrm{obs}\sim10\,000$\,y of the apsidal motion. From the orbital and fundamental stellar parameters obtained from our complex analysis we calculated the theoretically expected apsidal motion period (see Sect.\,\ref{lcanalysis}), and found to be $P_\mathrm{apse}^\mathrm{theo}\sim34\,000$\,y (see Table\,\ref{Tab:syntheticfit}). The insignificant length of 4 years of observations, compared to the ten-thousand-year-long period, could be attributed to be the main cause of the difference. We nevertheless cannot exclude the possibility of perturbations by a third star, which produces in a faster apsidal advance rate. A similar scenario has been detected in several {\it Kepler}-discovered hierarchical triple stellar systems (Borkovits et al., 2014, in preparation).  

\section[]{Spectroscopy}
\subsection{\label{Subsect.:fundparam}Fundamental parameters}

To determine the fundamental parameters, we used the two spectra taken at KPNO in 2011 and co-added to produce higher signal-to-noise ratio spectrum. We chose these two spectra because they cover large wavelength range and they have the highest signal-to-noise ratios among the spectra we had. We used the fitting recipe described in Shporer et al. (2011) based on crosscorrelating model spectra  by \citep{mun05} in the wavelength range of $5000 - 6400$\,\AA. This is a two-step method that first fits $T_\mathrm{eff}$, $\log g$ and $v_\mathrm{rot}\sin i_\mathrm{rot}$ assuming solar metallicity, and then accepting the effective temperature, the metallicity is refitted together with $\log g$ and $v_\mathrm{rot}\sin i_\mathrm{rot}$. This iterative method is stable in the high temperature range ($>7000$\,K) where $T_\mathrm{eff}$ and $[Fe/H]$ are significantly correlated. We found a preliminary solution of $T_\mathrm{eff}=7400 \pm 150$\,K, $\log g=3.5 \pm 0.3$, $v_\mathrm{rot}\sin i_\mathrm{rot}=100 \pm 10$\,km\,s$^{-1}$ and $[M/H]=-0.5 \pm 0.3$. This preliminary result was re-iterated by combining spectroscopic and photometric data. The most stable parameter is $v_\mathrm{rot}\sin i_\mathrm{rot}$, since its value is practically independent of the other three. This solution is also in good agreement with those in the Kepler Input Catalogue, $T_\mathrm{eff}=7500$ K, $\log g = 3.5$ and $[Fe/H]=-0.08$ \citep{bro11}.

Rapid rotation causes significant gravity darkening. According to the von Zeipel law \citep{zei24}, there is a temperature gradient reaching almost $1000$\,K on the surface of the primary. The effect of this gradient on the spectrum cannot be handled because we do not know the aspect angle of the spin axis. We think that this temperature gradient is the most important source of systematics in spectral modeling, and therefore the internal errors of fitting algorithms should be considered as indicative values.

Since the geometry is unknown, we fitted a complete set of unique spectra (instead of a weighted average of spectra to describe the temperature gradient), and also introduced stellar evolution tracks into the fitting procedure (Padova evolutionary tracks, \citealp{ber08,ber09}) to constrain the fit to components with compatible ages. We fitted jointly the stellar spectra, and observed the parameter set in the $T_{\rm eff}$ and $\log g$ isochrone. Moreover, we involved the second component in the fit, since its mass function, relative temperature and relative radius had been constrained from the light curves with acceptable precision at this stage of fitting. 

We searched for a solution that satisfied all the following criteria:

\begin{itemize}
\item The model is consistent with the measured KPNO spectra, according to a standard {$\chi^2$} analysis;
\item The model describes a valid position in the $T_\mathrm{eff}$ -- $\log g$ evolutionary track;
\item The model produces a secondary component which is also consistent with a valid position in the evolutionary track and has a similar age to that of the primary.
\end{itemize}

This iteration stabilized $T_\mathrm{eff}$ around $7400-7700$\,K, suggested a $\log g$ between $3.75-4.25$ depending on the age (which is larger than the fit of the spectrum alone), and also confirmed a slightly low metallicity (around $-0.5$). It is worth noting that the criterion of both stars having compatible ages confined the joint parameter set significantly, and resolved much of the known degeneracies of fitting a single spectrum. The new parameter set is consistent with a main sequence $\gamma$\,Dor star with rapid rotation and a fairly young age (see Table\,\ref{fundpar} for the determined parameters). The determined models within the confidence volume formed the allowed parameter set of the detailed light curve modelling (Sect.\ \ref{lcanalysis}) and describes all spectroscopic and photometric data well. In section 4.2 and 4.3, we will repeat the spectral analysis with disentangling. Although these two methods are based on partly differing input data and different data processing, the resulting stellar models are satisfactory compatible with each other, confirming the validity of the derived stellar parameters.

\begin{table}   
\begin{center}
 \caption{\label{fundpar} Fundamental parameters of the main component of HD\,183648 system adopted from spectrum fitting. }  
\begin{tabular}{lcc}  
\hline   
Parameter & Value & Error \\ 
\hline  
T$_{\rm eff}$ (K) & 7500 & 150  \\
$\log{g}$ (dex) & 4.0 & 0.25 \\
$\rm [M/H]$ (dex) & -0.5 & 0.3 \\
$v \sin{i}$ (km/s) & 100 &  10  \\
Age (Gyr)          & 0.9 & 0.2 \\
\hline   
\end{tabular} 
\end{center}  
\end{table}

\begin{figure}
\includegraphics[bb=55 195 431 481,width=84mm]{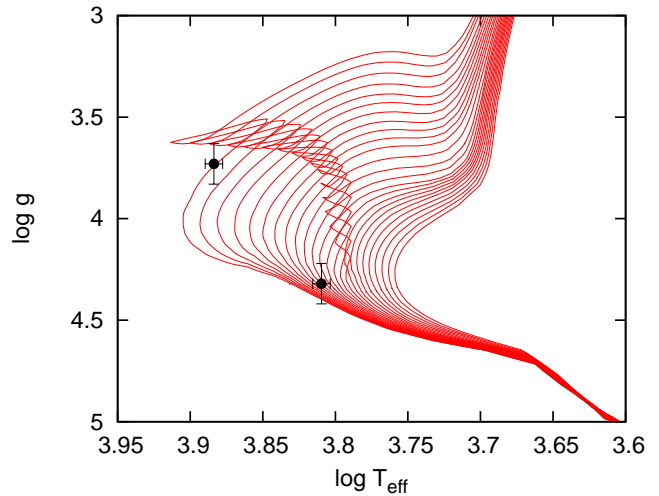}
\caption{The position of the two components on the T$_{\rm eff}$ -- $\log{g}$ evolution tracks with [Fe/H]\,=\,-0.5 metallicity. }
\label{Fig:evoltrack}
\end{figure}

\subsection{\label{Subsect:radvelstudy}Radial velocity study}

It has been usual since the epochal work of \citet{wilson79}, that radial velocity (RV) curves and photometric light curves are analysed simultaneously to obtain a combined solution. However, in the present situation we carried out these studies partially independently. The main reasons are as follows: while it is generally said that $e\cos\omega$ is very robustly determined by the light curve, this robustness is chiefly due to the timing, and not from any other parts of the light curve. On the other hand, for small eccentricities, the light curve itself has little dependence on $e\sin\omega$, which is better determined by the radial velocity data. These facts are especially valid for this present low-eccentricity system, where the out-of-eclipse parts of the light curve are strongly modulated by pulsations, which cannot be disentangled satisfactorily from the possibly anomalous ellipsoidal variations (see Sect.\,\ref{lcanalysis}). Therefore, we decided to obtain eccentricity ($e$) and argument of periastron ($\omega$) from the combination of iterative RV and eclipse timing solutions, and then to keep them fixed until the final refinement of the light curve solution. Note that other parameters of the spectroscopic and radial velocity solution (e.g., the spectroscopic mass function, and $v_\mathrm{rot}\sin i_\mathrm{rot}$) were also included in the light curve solution by constraining certain parameters; details of this are given below in Sect.\,\ref{lcanalysis}. 

The RV analysis was carried out iteratively combined with the ETV analysis. For the first run we used all the available radial velocity points (Table\,\ref{Tab:radveldata}). In this preliminary stage the systemic velocity $V_\gamma$, and the five usual orbital elements were adjusted by a Levenberg-Marquardt algorithm based non-linear least-squares fit, while the orbital period was kept fixed on the period determined in Sect.\,\ref{Sect:ETV}. Then, to check whether the period change that was detected in the ETV analysis manifests itself in the radial velocity curve as a variation in the systemic $V_\gamma$ velocity, an additional parameter, $\dot{V}_\gamma$, was also adjusted in an alternative run. 

\begin{figure}
\includegraphics[width=84mm]{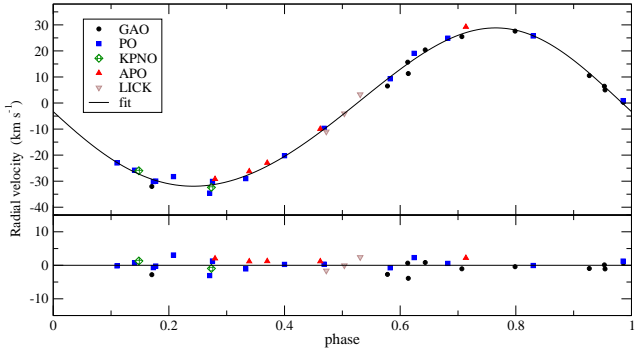}
 \caption{The observed radial velocties and the best fit radial velocity solution for the $\dot{V}_\gamma\equiv0$ case (solid line). The residuals of the fit are shown in the lower panel.}
 \label{Fig:rvsol}
\end{figure}

As a next step, the resulted eccentricity was used to refine the ETV solution, as discussed in Sect.\,\ref{Sect:ETV}. In this way we obtained refined $e$ and $\omega$ parameters that were consistent with the previous RV results, but had substantially smaller formal errors. Finally, we fixed the eccentricity to its ETV-fit value, and reiterated the RV fits. In these runs, although the argument of periastron ($\omega$) kept its large formal error of $>10\degr$, it converged to a value differing only by $\sim1\degr$ from the ETV solution. Our results are plotted in Fig.\,\ref{Fig:rvsol} ($\dot{V_\gamma}\equiv0$ solution), and listed in Table\,\ref{tab: RVresult}. In the last rows we give some additional, derived quantities. As one can see, the apparent period variations ($\Delta P$), which were calculated from $\dot{V}_\gamma$, are slightly higher, yet agree with the result obtained from the ETV analysis.

In Fig.\,\ref{Fig:rvsol} (lower panel) the residual velocity data are also plotted. As one can observe, these values exceed the estimated observational uncertainties for most of the data points. In order to investigate whether these deviations come from stellar pulsation, and/or instrumental effects, we performed a test. We checked whether the residuals of the radial velocity data show correlations with instrumental parameters such as spectral resolution and S/N, or are more likely of non-instrumental origin. The S/N was calculated near the blue wing of the H$\alpha$ line, between  640 and 645~nm, where the spectrum is nearly featureless. We estimated the continuum S/N levels to be between 40--300. The scatter of the radial velocity residuals did not exhibit a correlation neither with S/N nor the resolution of the spectra. The median absolute deviation of the residuals was $700$\,m\,s$^{-1}$ for the $\dot{V}_\gamma\neq0$ solution (i.e. observation -- model, assuming a long-term component to describe the effect of the assumed third companion), regardless of the instrumental parameters. Moreover, the residuals did not show any periodicity, which could be related to the observed pulsation (see Sect.\,\ref{lcanalysis}). Thus, the origin of this wobbling remains unexplained. Nevertheless, since the full amplitude of the radial velocity curve is over $50$\,km\,s$^{-1}$, the velocity wobbling is under 2\% and does not influence the dynamical analysis. 

\begin{table}
 \caption{Results of radial velocity solutions, and some derived parameters (probable errors in the last digits). Note, reference epoch ($T_0$) and period $P_\mathrm{orb}$ were kept fixed.}
 \label{tab: RVresult}
 \begin{tabular}{@{}lcc}
 \hline
  Parameters & $\dot{V}_\gamma\equiv0$ & $\dot{V}_\gamma$ adjusted \\
  \hline
  $T_0$ (BJD) & \multicolumn{2}{c}{$2\,454\,966.8687$}\\
  $P_\mathrm{orb}$ (d) & \multicolumn{2}{c}{$31.97312$}\\
  \hline
  $(V_\gamma)_0$ (km\,s$^{-1}$) & $-2.6(3)$ & $-6.3(17)$ \\
  $\dot{V}_\gamma$ (km\,s$^{-1}$/cycle) & $0(-)$ & $0.104(46)$ \\
  \hline
  $a_1\sin i$ (R$_{\odot}$) & $19.08(24)$ & $19.04(24)$ \\
  $e$ & $0.050(13) $  & $0.048(13)$ \\
  $\omega$ ($\degr$) & $43.9(129)$ & $38.4(137)$ \\
  $M_0$ ($\degr$)   & $44.2(129)$ & $49.0(136)$ \\
  \hline
  $\tau$ (BJD)& $2\,454\,962.9(11)$ & $2\,454962.5(12)$ \\
  $K_1$ (km\,s$^{-1}$) & $30.25(39)$ & $30.18(36)$ \\
  $f(m_2)$ (M$_{\odot}$) & $0.0911(35)$ & $0.0906(35)$ \\
  $\Delta P\mathrm{~(d/cycle)}$ & $-$ & $1.1(5)\times10^{-5}$ \\
  \hline
 \end{tabular}

\end{table}

\subsection{Detection of the secondary component and dynamic masses}                                           
\label{sec: spd}

The spectra of the faint secondary component was not detected in the crosscorrelation function (CCF) (see Sect.\,\ref{Subsect.:fundparam}). Therefore, the direct dynamical determination of the masses of the components has not been possible. Furthermore, the secondary's spectral properties have also remained unclassified. None of this information is crucial for the complex analysis of the system, as neither the orbital nor the light curve solutions are dependent on the stellar masses. Moreover, the less than 5\% contribution of the secondary's light to the total flux of the system suggests, that the composite spectra, and therefore the CCF solution of the primary is only weakly affected by the contribution of the secondary. However, from astrophysical point of view, stellar masses are the most important parameters. Therefore, in order to obtain dynamical masses and additional information on the secondary we made additional efforts to separate the signal of the secondary from the composite spectra.

The method of spectral disentangling ({\sc spd}) enables isolation of the indvidual component spectra simultaneously with the determination of the optimal set of orbital elements \citep{simonsturm94,hadrava95}. A time-series of the spectra are needed spread along the orbital cycle. Faint components are detected by {\sc spd} in the high-resolution spectra \citep[c.f.][]{pavlovskietal09,lehmannetal13,tkachenkoetal14} but a good phase coverage and a high S/N are needed. This is an important feature of {\sc spd} since the disentangled spectra are effectively co-added from the original observed spectra and so have a higher S/N. 

Our spectroscopic data sets are of different spectral resolutions and S/N (Sect.\,\ref{Subsect:Spectroscopy}). Several spectra per night were usually obtained at Piszk\'estet\H o and Gothard Observatories and we stacked them to enhanced S/N. Still some of these stacked spectra suffered from low S/N and were not used in {\sc spd}. We decided to omit these spectra, and after the selection we dealt with 16 spectra suitable for {\sc spd} (the observed spectra used in {\sc spd} are indicated in Table~\ref{Tab:radveldata} by asterisk). Fortunately, selected spectra cover a complete orbital cycle and hence fulfill a prerequisite for a stable disentangling. Because of different resolution we re-sampled all spectra to medium resolution of GAO spectra. We assigned the weights according to the S/N, and an initial spectral resolution.

\begin{figure*}
\includegraphics[width=18cm]{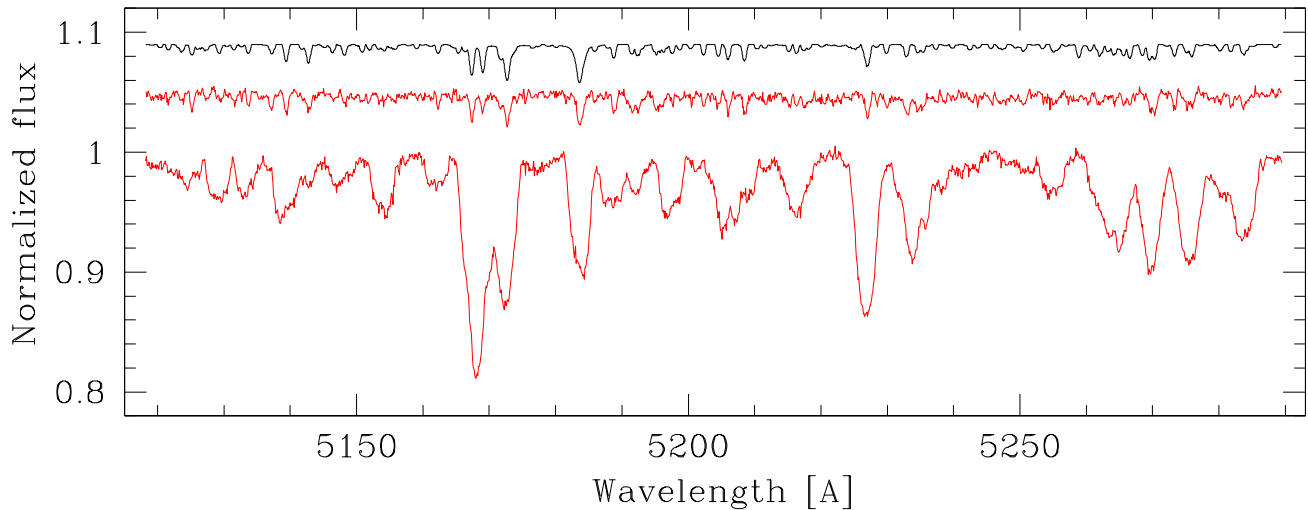}
\caption{Disentangled spectra of the components in the binary system HD~183648 (lower two spectra) in the spectral region centered on the \ion{Mg}{i} triplet $\lambda\lambda 5167-5184$ {\AA}. For comparison, the top curve is a synthetic spectrum of a star with atmospheric parameters of the secondary (c.f.~Table\,\ref{Tab:syntheticfit}), rotationally broadened with $v_\mathrm{rot}\sin i_\mathrm{rot} = 26$ \,km\,s$^{-1}$\ and diluted by a factor of 20.}
\label{figDisent}
\end{figure*}

The code {\sc FDBinary} \citep{ilijicetal04} which implements disentangling in the Fourier domain \citep{hadrava95} was used to  perform {\sc spd} in spectral regions centred on the \ion{Mg}{i} triplet, $\lambda\lambda 5167-5184$ {\AA}, covering about 200\,{\AA}. In these calculations eccentricity, $e$, and the argument of periastron, $\omega$, were set fixed, as these orbital elements were better constrained from the combined RV+ETV analysis (sects.\,\ref{Subsect:radvelstudy}\,\ref{Sect:ETV}). Then the orbital solution obtained by {\sc spd} yielded velocity semi-amplitudes of $K_1 = 34.4\pm1.1$ \,km\,s$^{-1}$ and $K_2 = 62.3\pm1.6$ \,km\,s$^{-1}$, and thus a mass ratio $q = 0.552\pm0.023$. The quoted errors for the semi-amplitudes derived by {\sc spd} were calculated by the `jackknife' method \citep[c.f.][]{pavlovskisouthworth09}. A comparison with the single-lined RV study reveals that {\sc spd} resulted an $\sim13\%$ larger primary semi-amplitude, and consequently, via the spectroscopic mass-function, a higher mass-ratio. The discrepancy might come from two reasons, either (i) the unresolved light contamination of the secondary's spectra to the primary's spectral lines in CCF measurements, which acts to reduce the semi-amplitude of the primary RV curve, or (ii) the same effect which causes the radial velocity residual wobbling, discussed in Sect.\,\ref{Subsect:radvelstudy}, resulting in slight spectral variations and therefore, slightly biases the disentangling. Note, a thorough analysis of different CCF and disentangling methods were carried out in \citet{southworthclausen07}, who also found that {\sc spd} gives higher (and more reliable) semi-amplitudes, especially when the spectra were affected by line-blending. 
A portion of disentangled spectral region is shown in Fig.\,\ref{figDisent}. The spectrum of a mid-F type secondary component is clearly revealed as could be judged by comparison with the syntethic spectrum for its atmospheric parameters and diluted by the factor of 20 to mimic its contribution to the total light of the system. As shown in Fig.\,\ref{figDisent} {\sc spd} was performed in the `separation' mode, and with the known light ratio (from the light-curve analysis). These separated spectra are renormalised to the continua of the respective components \citep{pavlovskihensberge05} for the further detailed spectroscopic analysis. For this latter process we fixed $\log g$s and light factors on the values found from the combined detailed CCF-spectroscopic and light curve analyses (see Table\,\ref{Tab:syntheticfit}), and fitted only temperatures and projected rotational velocities. Then our analysis resulted in $T_\mathrm{eff1}=7510\pm90$\,K, $\left(v_\mathrm{rot}\sin i_\mathrm{rot}\right)_1=104.2\pm1.5$\,kms$^{-1}$ and $T_\mathrm{eff2}=6490\pm140$\,K, $\left(v_\mathrm{rot}\sin i_\mathrm{rot}\right)_2=26.0\pm2.4$\,kms$^{-1}$ for the primary and the secondary, respectively. Therefore, the effective temperatures were found to be in accordance with the results of the combined analysis within their errors (see Table\,\ref{Tab:syntheticfit}). The main significance of this result lies in the substantially reduced uncertainty of the primary's projected rotational velocity, and the determination of the same parameter for the secondary component. We note also that the temperature ratio obtained from disentangled spectra of the components was found to be $0.864\pm0.021$ in contrast to the photometrically found value $0.843\pm0.017$, with a difference within the uncertainty limit. This is an additional interdependent verification of the results obtained in different manners.

\section[]{Light curve analysis}
\label{lcanalysis}

\begin{figure*}
\includegraphics[width=0.95\linewidth]{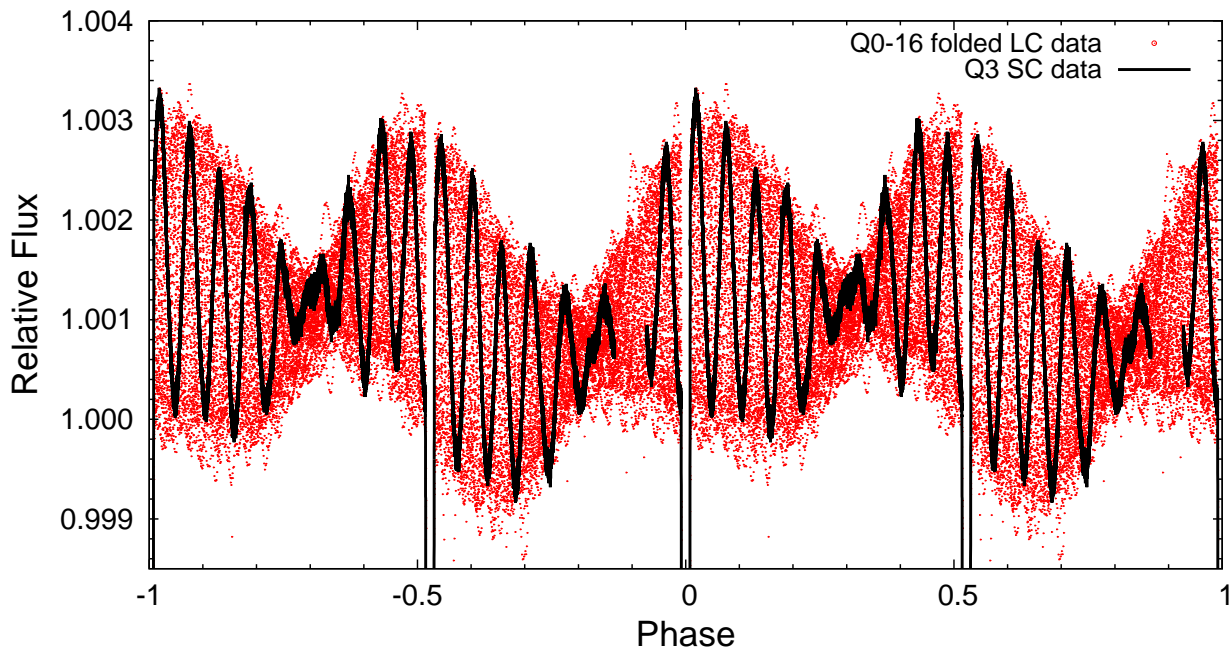}
\caption{Folded light curve of HD\,183648 around the maximum, out of eclipse light level (the eclipses are therefore off scale). The folded total Q0 -- Q16 long cadence light curve illustrates well that the nodes of the beating phenomenon remained at constant orbital phases over the whole 4-y data set (red data). Black data represent the Q3.2 (only) short cadence observations, which show the pulsational pattern over one orbital cycle.}
\label{Fig:envelope}
\end{figure*}

The {\it Kepler} light curve reveals at least three different features. The most prominent pattern shows that HD\,183648 is a relatively long-period ($P_\mathrm{orb}=31.973$\,d) eclipsing binary on an eccentric orbit. The light curve also shows pulsations with periods near 1.78\,d. Moreover, the amplitudes of these pulsations shows an obvious beat phenomenon with a period that is equal to half of the orbital period. As a consequence, the maxima and minima of the envelope of the pulsation occur at the same orbital phases during the whole 4-year observational interval. Furthermore, another sinusoidal light variation is also observable with a period equal to half-orbital period, and phased in such a way that the maximum brightnesses occur near orbital phases $0.0$ and $0.5$ (i.e., near the eclipses). Therefore, this enigmatic variation looks like an ``inverse'' ellipsoidal effect, or resembles a reflection or irradiation effect, although its high amplitude clearly excludes this latter explanation. An additional sinusoidal brightness variation with a period equal to the orbital period is also observable, however, as it will be shown later, this latter feature well can be explained by Doppler boosting. All these light curve features are illustrated in Fig.\,\ref{Fig:envelope}. 

In order to obtain a physically correct binary star model, the different properties of these complex light curve variations were disentangled. As simultaneous eclipsing binary and pulsation modelling methods and programs are not available yet, we followed an iterative procedure, similar to that which was described and applied in the papers of \citet{mac13} and \citet{debosscheretal13}. This method is based on the rectification of the light curve with an iterative separation and then, removal of the other light curve variations from the eclipsing binary features, by the use of Fourier space, obtaining an approximately pure eclipsing binary light curve, which can then be fitted by a light curve fitting algorithm. Following the removal of this solution from the original curve, a more accurate pulsation pattern can be obtained. This method can lead to an improved pulsation model that is then removed from the original light curve to obtain a more improved eclipsing binary light curve. In the present situation, however, the presence of the exactly half orbital period extra variation provides a slight complication, as it covers the possibly small ``normal'' ellipsoidal effect, and modifies eclipse depths and shapes coherently in phase. Note that \citet{sou11} explained the unphysical outputs of their light curve solution for KIC\,10661783 with such an effect. Fortunately, however, the amplitude of this variation is less than 1\,mmag, and consequently, it has only minor effect on the eclipses.

\begin{figure*}
\includegraphics[width=84mm]{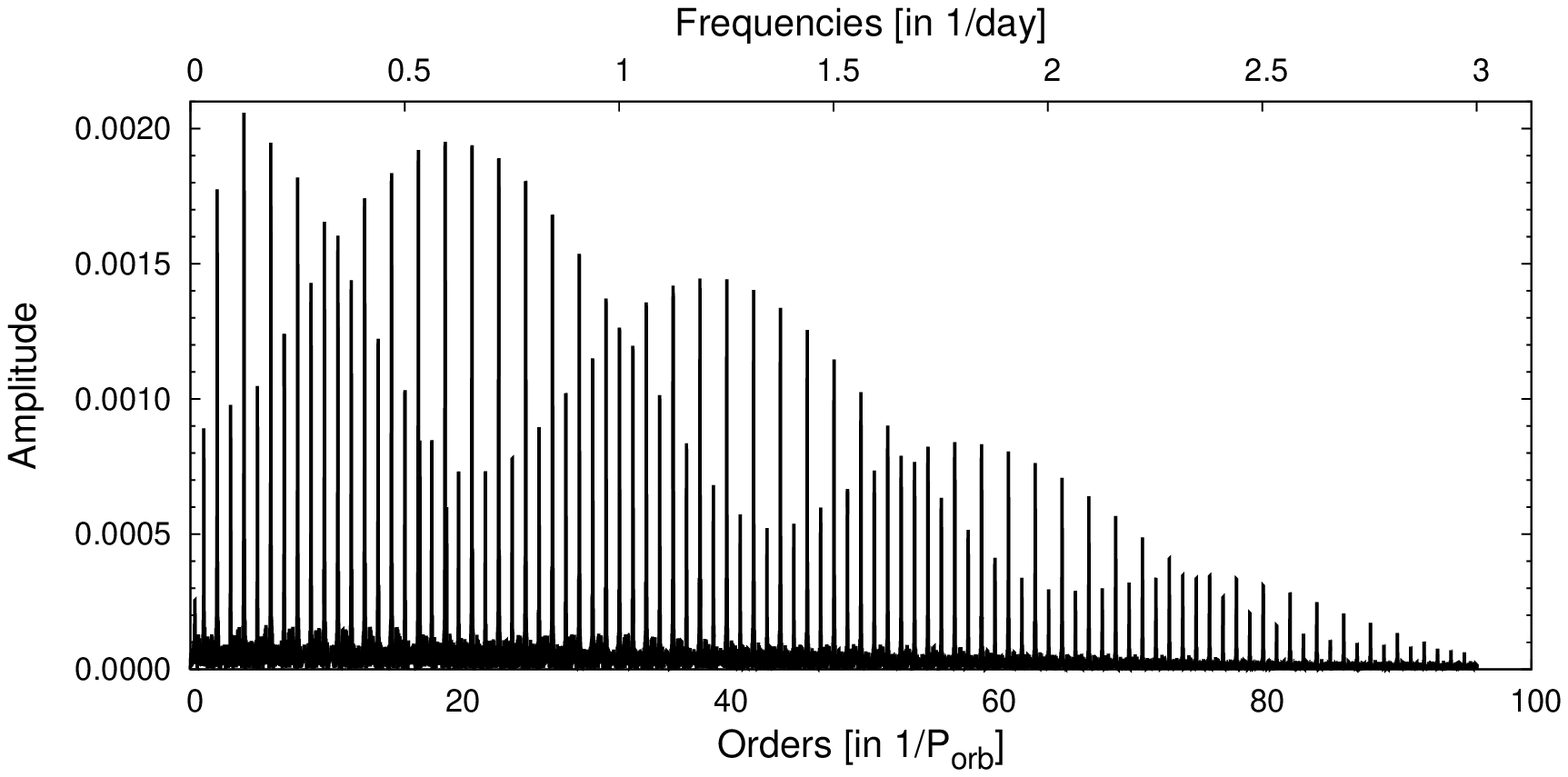}\includegraphics[width=84mm]{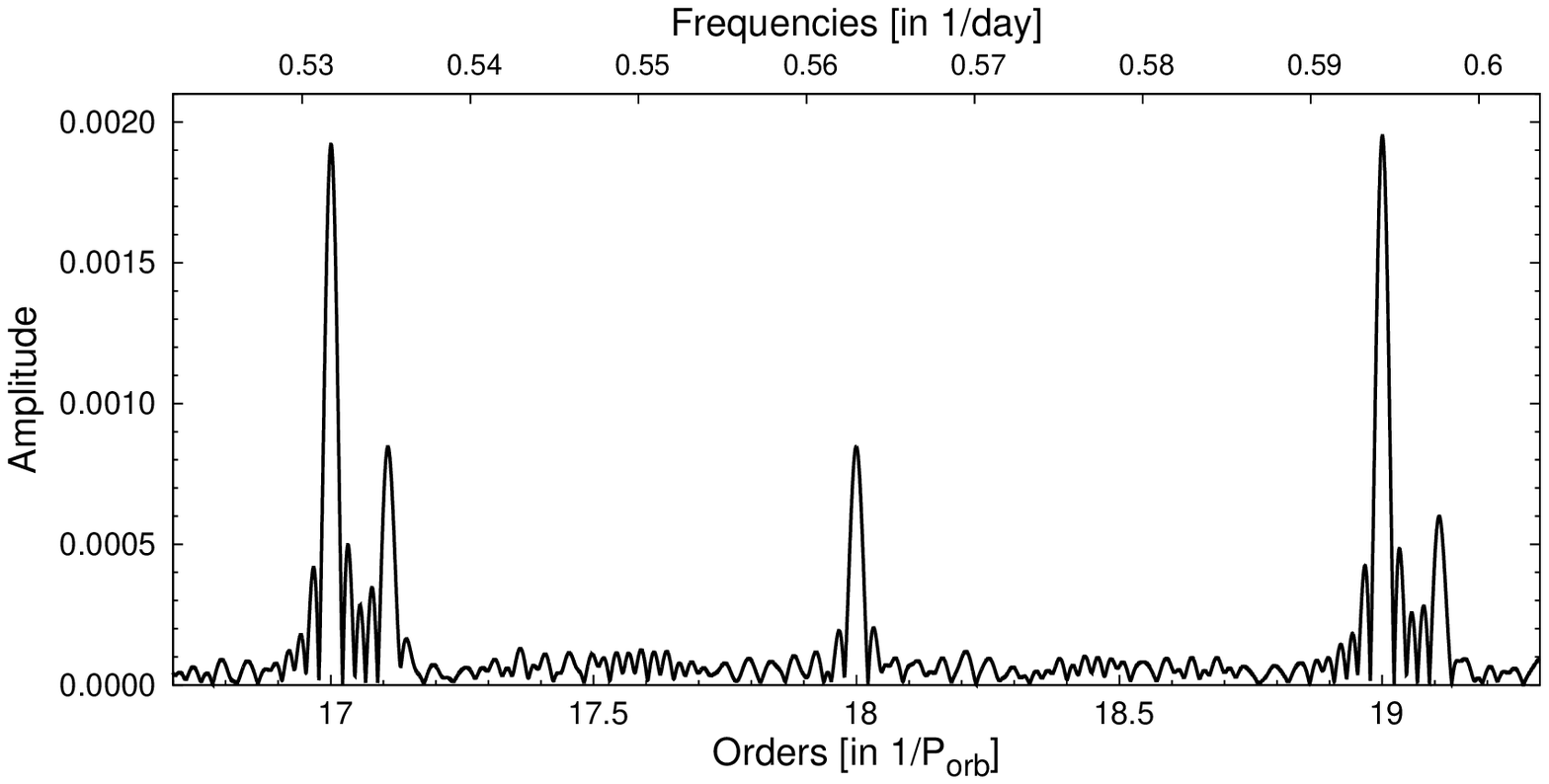}
\caption{Discrete Fourier Transform amplitude spectra of the detrended Q0 -- Q16 LC light curve of HD\,183648. Left panel: The complete amplitude spectrum up to $f=3.0\,\mathrm{d}^{-1}\approx95f_\mathrm{orb}$. The harmonics of the orbital frequency dominate almost exclusively.  Right panel: The two pulsational frequencies in the vicinity of frequencies $17f_\mathrm{orb}$ and $19f_\mathrm{orb}$ are separated exactly by $2f_\mathrm{orb}$.}
 \label{Fig:DFTraw}
\end{figure*}

For the initial disentangling of the pulsation and eclipse patterns, we calculated the Discrete Fourier Transform amplitude spectrum of the raw data. As one can see in Fig.\,\ref{Fig:DFTraw}, a very regular spectrum was obtained which contains harmonics of the orbital frequency almost exclusively. There are two main pulsation peaks separated equally in frequency from $17f_\mathrm{orb}$ and $19f_\mathrm{orb}$.

After obtaining these two pulsation frequencies, we carried out a four-frequency linear least-squares fit  
on the out-of-eclipse parts of the raw, detrended Q0 -- Q16 LC light curve. We fit not only the two dominant pulsation frequencies, but also, in accordance with the additional light curve features mentioned above, the frequencies $f_\mathrm{orb}$, $2f_\mathrm{orb}$, too. Then we removed this least-squares solution from the raw curve. It is evident that we possibly removed the ellipsoidal, reflection and Doppler-boosting effects. However, from the preliminary system characteristics and light curve properties, we expected only minor (if any) contributions from ellipsoidal and reflection effects. Regarding Doppler-boosting, the version of the PHOEBE software package \citep{prsazwitter05} that was used in this preliminary stage does not model it. 

The initial values of the fundamental parameters for the primary star were taken from the spectroscopic results, while the orbital elements were taken from the preliminary radial velocity and ETV analyses. 
The differential correction part of the PHOEBE analysis was applied for three different datasets, namely, for the pulsation-removed (i) Q0 LC data, (ii) Q3.2 SC data and finally, (iii) the binned, averaged Q0 -- Q16 LC data.

After the removal of the PHOEBE solution an improved pulsation model was calculated and subtracted in a similar manner. Then, after reaching a quick convergence of this iterative method, we made a final parameter refinement with our own {\sc lightcurvefactory} light curve synthesis program \citep{borkovitsetal13}. As a recent improvement, a linear least-squares based multi-frequency Fourier polynomial fitting subroutine was also built into the code, which made it possible to fit quasi-simultaneously both the eclipsing binary and the pulsation models internally, at every step. Note that such a combination has only practical and time saving advantages, but remains an unphysical solution for the combined investigation of pulsation and binarity effects, and hence suffers from all the disadvantages that were discussed in \citet{wilsonvanhamme10}. We used five frequencies for the Fourier fitting procedure, namely, the four higher amplitude pulsation frequencies (see Sect.\,\ref{Subsect:freqsearch}), and $2f_\mathrm{orb}$.
As our program also takes Doppler-boosting into account, we removed the $f_\mathrm{orb}$ frequency component from the Fourier fitting. Furthermore, in this refinement process, the rotation synchronization parameter was no longer kept fixed, but was constrained according to the spectroscopically determined values of $v_\mathrm{rot}\sin{i}_\mathrm{rot}$ (assuming that $i_\mathrm{rot}=i_\mathrm{orb}$).

For this next combined refinement, the complete, previously detrended, unaveraged, unbinned Q0--Q16 long cadence light curve was used. This curve contains $64\,528$ data points. In order to reduce computing time, we switched off the computation of the reflection effect (which is by far the most time-consuming part of the calculations), and used four-times coarser stellar grids in the out-of-eclipse phases. A test verified that reflection/irradiation affected the light curve around the secondary minima (where it reaches its maximum), at an insignificant 10\,ppm level, while the coarser grid was found to have no systematic effect on the goodness of a given parameter set, but resulted in a somewhat higher $\chi^2$ value due to the noisier synthesis curve. Naturally, reaching a convergent solution, the final light curves and residuals were calculated with reflection and finer grids.

\begin{figure}
\includegraphics[width=84mm]{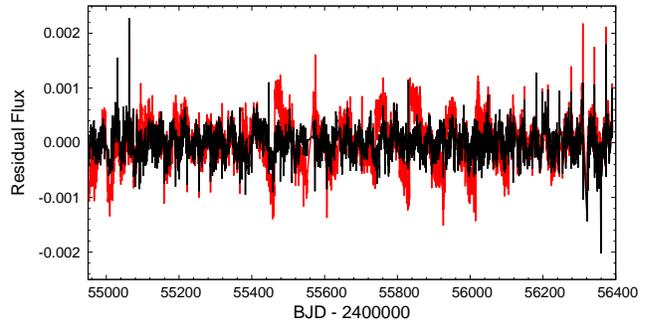}
 \caption{The residuals of the combined eclipsing + 5-frequency pulsation solution before (red) and after (black) of the final detrending. (See text for details.)}
 \label{Fig:resbeforeafter}
\end{figure}

Calculating the refined solution in this manner, the residual curve revealed that there were systematic discrepancies in certain Quarters. This manifested itself in non-zero average slopes of some quarterly data. Although it cannot be excluded that these effects have physical origins (e.g., longer time-scale brightening or fading of the system, or a residual effect of the Doppler-boosting caused by the {\it Kepler} spacecraft's motion), from our point of view, they represent additional, systematic noise which should be removed. Therefore, we fitted the residual data with first order polynomials, individually for each quarter, and by the use of them, we detrended again the previously used Q0 -- Q16 LC data. Then, we repeated only the last, refining part of our analysis. The residuals of the combined eclipsing and 5-frequency pulsation curve before and after this final detrending are plotted in Fig.\,\ref{Fig:resbeforeafter}.  

\begin{table}
\caption{Stellar and orbital parameters derived from the combined radial velocity, eclipsing light curve and ETV analysis.}
 \label{Tab:syntheticfit}
 \begin{tabular}{@{}lll}
  \hline
\multicolumn{3}{c}{orbital parameters} \\
\hline
  $P_\rmn{orb}$ (days) & \multicolumn{2}{c}{$31.97325\pm0.00002$} \\
  $T_\rmn{MIN I}$ (BJD) & \multicolumn{2}{c}{$2\,454\,966.8687\pm0.0002$} \\
  $a$ (R$_\odot$) & \multicolumn{2}{c}{$61.08\pm1.27$} \\
  $e$ & \multicolumn{2}{c}{$0.0477\pm0.0020$}  \\
  $\omega$ ($\degr$)& \multicolumn{2}{c}{$40.08\pm0.08$}  \\ 
  $i$ ($\degr$) & \multicolumn{2}{c}{$87.32\pm0.15$}  \\
  $\tau$ (BJD) & \multicolumn{2}{c}{$2\,454\,962.798\pm0.025$} \\
  $q$             & \multicolumn{2}{c}{$0.5523\pm0.0226$} \\
  $P_\rmn{apse}^\rmn{obs}$ (years) & \multicolumn{2}{c}{$10\,400\pm3\,000$} \\
\hline
\multicolumn{3}{c}{theoretically derived orbital parameters} \\
\hline
 $P_\rmn{apse}^\rmn{theo}$ (years) & \multicolumn{2}{c}{$34\,200\pm2\,000$} \\
 $\dot\omega_\mathrm{rel}^\rmn{theo}$ (arcsec/$P_\mathrm{orb}$) & \multicolumn{2}{c}{$2.547\pm0.167$} \\
 $\dot\omega_\mathrm{cl}^\rmn{theo}$  (arcsec/$P_\mathrm{orb}$) & \multicolumn{2}{c}{$0.772\pm0.103$} \\
  \hline  
\multicolumn{3}{c}{stellar parameters} \\
\hline
   & Primary & Secondary \\
  \hline
\multicolumn{3}{c}{fractional radii} \\
  \hline
 $r_\rmn{pole}$  & $0.05240\pm0.00010$ & $0.01810\pm0.00020$ \\
 $r_\rmn{side}$  & $0.05475$ & $0.01813$ \\
 $r_\rmn{point}$ & $0.05476$ & $0.01813$ \\
 $r_\rmn{back}$  & $0.05476$ & $0.01813$ \\
 \hline
 \multicolumn{3}{c}{absolute stellar parameters} \\
  \hline 
 $M$ (M$_\odot$) & $1.93\pm0.12$ & $1.06\pm0.08$ \\
 $R$ (R$_\odot$) & $3.30\pm0.07$ & $1.11\pm0.03$ \\
 $T_\mathrm{eff}$ (K)& $7650\pm100$ & $6450\pm100$ \\
 $L$ (L$_\odot$) & $32.88\pm0.20$ & $1.87\pm0.12$ \\
 $\log g$ (dex) & $3.71\pm0.03$ & $4.38\pm0.04$ \\
 $P_\mathrm{rot}$ (days) & $1.60\pm0.04$ & $2.15\pm0.21$ \\
 \hline
 \end{tabular}
\end{table}

\begin{table}
 \caption{Model-dependent (readjusted) and fixed parameters}
 \label{Tab:syntheticfix}
 \begin{tabular}{@{}lll}
  \hline
  Parameter & Primary  & Secondary \\
\hline
Linear limb darkening (bolometric)     & $0.6658$ & $0.6658$  \\
Logarithmic limb darkening (bol.)      & $0.2493$ & $0.1701$  \\
Linear limb darkening (monochrom.)     & $0.6121$ & $0.6191$  \\
Logarithmic limb darkening (mono.)     & $0.2350$ & $0.1799$  \\
First apsidal motion constant ($k_2$)  & $0.0020$ & $0.0080$  \\
Second apsidal motion constant ($k_3$) & $-$      & $0.0020$  \\
Bolometric albedo                      & $1.0$    & $0.6$     \\
Gravity darkening exponent             & $1.0$    & $0.32$    \\
\hline
\end{tabular}
\end{table}

\begin{figure*}
\includegraphics[width=168mm]{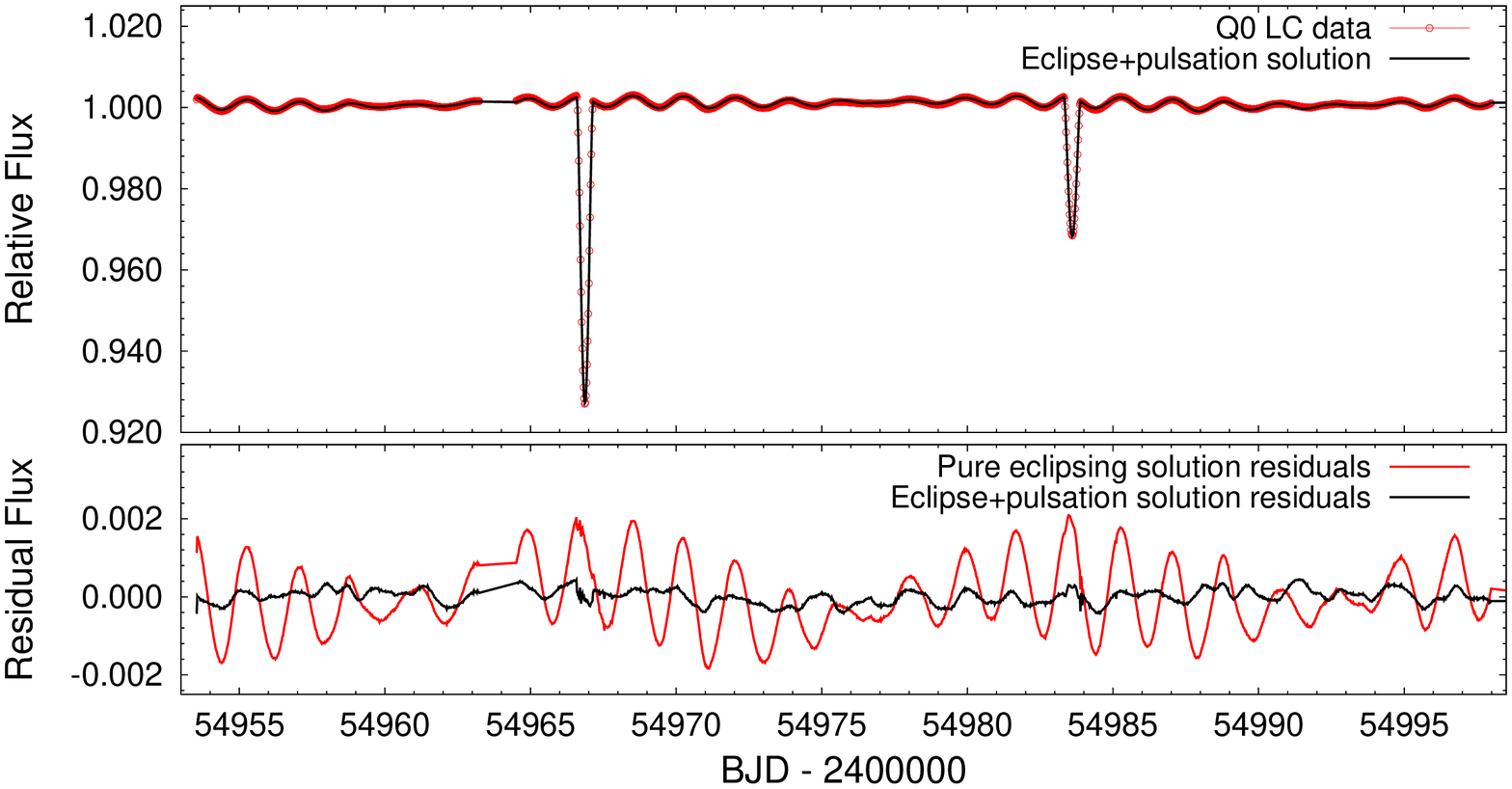}\\
\includegraphics[width=168mm]{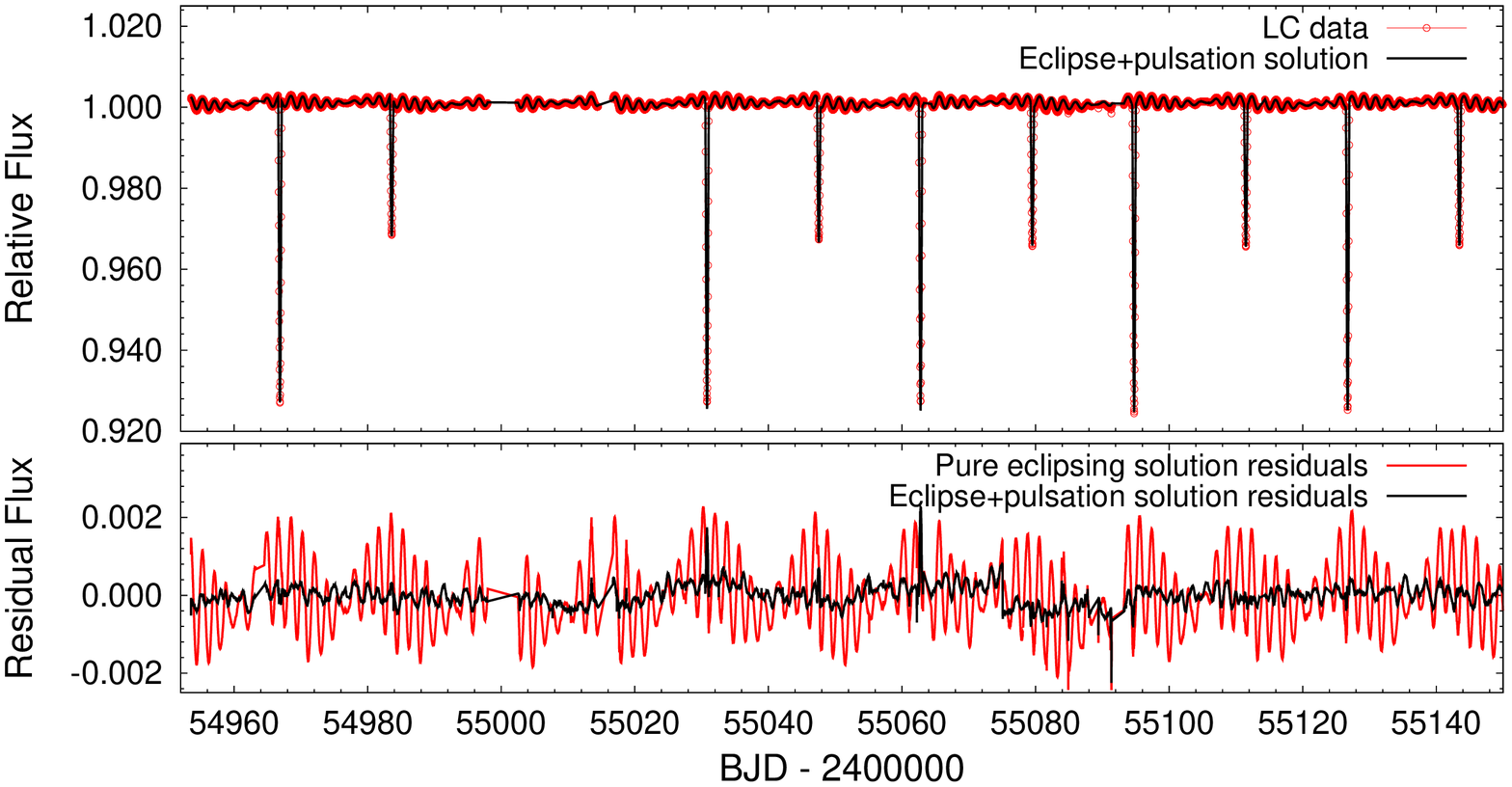}
 \caption{{\it Upper panels:} Parts of the final combined eclipse and 5-frequency pulsation light curve solution (black line) for the detrended Kepler LC data (red dots). {\it Bottom panels:} The residuals are for the combined solution curve (black), and its pure eclipsing contribution (red). The latter represents the pulsation component of the observed data, which is further analysed in Sect.\,\ref{Subsect:freqsearch}.}
 \label{Fig:Q0lcsolution}
\end{figure*}

\begin{figure}
\includegraphics[width=84mm]{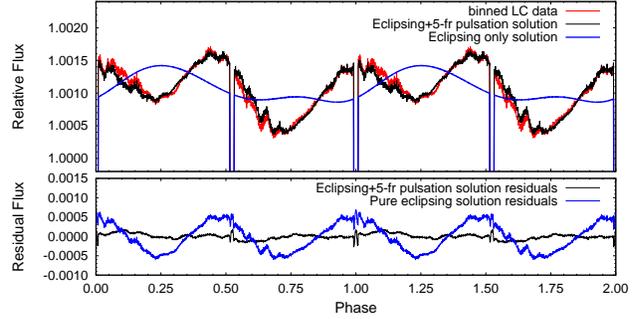}
\caption{{\it Upper panel:} The folded and binned out-of-eclipse section of the whole Q0 -- Q16 detrended LC data (red) together with the combined, simultaneous eclipse and 5-frequency pulsation solution (black) and with the pure eclipsing part of the same solution (blue). {\it Bottom panel:} The folded, binned residuals of the solutions above.}
 \label{Fig:lcsolution_folded}
\end{figure}

The final results of the combined eclipsing light and radial velocity curve analysis are listed in Tables\,\ref{Tab:syntheticfit} and \ref{Tab:syntheticfix}, and are shown in Fig.\,\ref{Fig:Q0lcsolution}. Furthermore, Fig.\,\ref{Fig:lcsolution_folded} gives details on the out-of-eclipse part of the folded and binned solution, which is plotted there both with and without the 5-frequency pulsation. The latter corresponds to the theoretical, pure eclipsing binary light curve solution, i.e. the sum of the ``normal'' ellipsoidal effect and Doppler boosting (blue curve in the upper panel). The regular half-orbital-period sinusoidal shape of the phased residuals of this theoretical curve (blue curve in the lower panel of Fig.\,\ref{Fig:lcsolution_folded}) (i.e., the absence of the brightness differences between the two quadratures) demonstrates clearly that the $f_\mathrm{orb}$ component is well described purely with Doppler boosting. This also gives independent evidence for the absence of significant third light in the light curve. If there were significant third light, the additional light contribution would reduce the observable amplitude of Doppler-boosting and, consequently, the theoretical fit would overestimate it. This result suggests that if there is a third companion, it should be probably a low-mass M dwarf star (see Sect.\,\ref{Sect:ETV}).

The oscillatory features of the residual curve will be discussed in the next section. Here we only comment on the small residual discrepancies during the two kinds of minima. What is surprising is not their presence (they occur commonly in the case of very accurate satellite light curves due to the incomplete physics included in the presently available models, see \citet{ham13} for a short discussion), but that their amplitudes do not exceed $300-500$ ppm in relative flux. Taking into account the irregular, and therefore incompletely modelled, ellipsoidal effect (to be discussed in the next Section), we are inclined to take the extraordinay goodness of our fit as a mere coincidence and not the outcome of a serendipitously found accurate physical model. 

In Table\,\ref{Tab:syntheticfit} we tabulate some derived quantities, such as the rotational period ($P_\mathrm{rot}$) of the two components, and the theoretical relativistic and classical tidal apsidal motion angular velocities. For this calculation the apsidal motion constants (listed in Table\,\ref{Tab:syntheticfix}) were taken from the tables of \citet{claretgimenez92}. Note that for the calculation of the tidally forced apsidal motion we used only the equilibrium tide model \citep{cowling38,sterne39}, and did not consider the dynamical contribution \citep[see, e.g.,][]{claretwillems02}. A proper calculation of the dynamical tides for the fast rotating primary is beyond the scope of the present paper. It should be stressed, however, that in the case of resonant tidal locking, the contribution of the dynamical tides may exceed the classical ones \citep{willemsclaret05}. This fact might offer an additional explanation for the discrepancy between the calculated apsidal advance rate and the observed one, which was examined previously in Sect.\,\ref{Sect:ETV}. The role of the fast rotation of the primary in the tidal oscillations will be discussed below in Sect.\,\ref{Subsect:tidaloscillation}.

The uncertainties of the parameters were determined with various methods. For the ETV and the radial velocity analysis, the errors given are mostly the formal errors of the differential correction procedures. It is well known, however, that these formal errors underestimate the real uncertainties due to the strongly degenerate nature of the eclipsing binary light curve modelling, with substantial correlations among the parameters, and should not be taken too seriously. Therefore we resorted to the more realistic estimations given by the final refinement to the light curve solution, which was essentially a Monte Carlo simulation. Our experiences are in accordance with those found by \citet{ham13} in a similar situation. Therefore we conclude that, despite the significant correlations, the light curve parameters are relatively well determined for this sort of detached {\it Kepler} binary with significantly deep eclipses.

\section{Oscillations and tidal effects}

\subsection{\label{Subsect:freqsearch}Frequency search}

\begin{figure*}
\includegraphics[width=168mm]{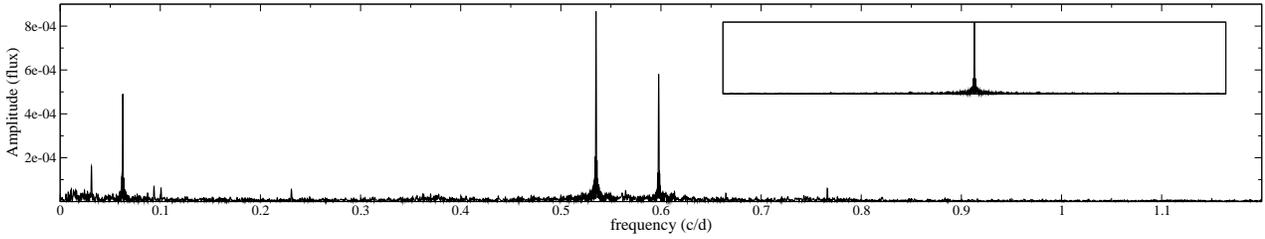}
\caption{\label{fourier} The pulsational amplitude spectrum of 17 quarters of long cadence data for HD\,183648. The insert shows the spectral window.}
\end{figure*}

After subtracting the eclipses, rotation and other binary related variations (see Sect.\,\ref{lcanalysis}), we analysed the remaining nearly continuous data set containing mainly the pulsations. For the period analysis we used {\sc period04} \citep{len05}, least-squares fitting of the parameters was also included and the signal-to-noise ratio ($S/N$) of each frequency was calculated following \citet{bre93}. The resulting significant peaks are listed in Table\,\ref{freqs}, while the Fourier spectrum is shown in Fig.\,\ref{fourier}.

We identified the two main pulsation frequencies at $F_{1}={\rm 0.535157 (1)\,d^{-1}}$ and $F_{2}={\rm0.597712(1)\,d^{-1}}$. The most intriguing result is that $F_{2}-F_{1}$ is exactly equal to $2 f_{orb}$ within $0.000003\,d^{-1}$, suggesting tidal origin. A further 6 statistically significant peaks were identified in the data. The $F_8$, $F_3$ and $F_4$ peaks represent the orbital frequency, and its second and third harmonics, respectively. While the less-significant $F_8$ frequency (i.e. the orbital frequency) might purely be the remnants of the light curve solution, the two higher harmonics are thought to be real, and will be discussed below. Furthermore, $F_{5}={\rm 0.766149\,d^{-1}}$ might also be interpreted as an independent oscillation frequency, which is supported by the fact that $F_{7}={\rm 0.230964\,d^{-1}}$ is equal to $F_{5}-F_{1}$ within $0.000028\,d^{-1}$. Finally, $F_{6}={\rm 0.100662\,d^{-1}}$ might be a remnant of the light curve fit or an instrumental effect. 

\begin{table*}   
\begin{center}
\caption{\label{freqs} The significant peaks of the period analysis. (Phases are calculated for periastron passage $\tau=2\,454\,962.798$.)}  
\begin{tabular}{lcccrc}  
\hline   
& Frequency  & Amplitude  & phase & S/N & Orbital  \\
 & (d$^{-1}$) &    ($\times 10^{-04}$ flux) &  (rad) & & solution \\
\hline   
F1$^*$& 0.535157(1) & 8.711(11) & -0.2200(13) & 88  &  $f_{1}$ \\
F2$^*$& 0.597711(1) & 5.880(11) & -2.6713(19) & 61  &  $f_{2}$ \\
F3$^*$& 0.062551(1) & 4.946(11) & -1.3473(22) & 40  &  $2 \cdot f_\mathrm{orb}$ or $f_{2}-f_{1}$ \\
F4& 0.093783(1) & 0.6805(112) &  2.0380(165) &  6  &  $3 \cdot f_\mathrm{orb}$ \\
F5$^*$& 0.766149(1) & 0.6267(112) &  1.9013(180) &  7  &  $f_{3}$  \\
F6& 0.100660(1) & 0.6199(112) & -2.9061(182) &  5  &   \\
F7$^*$& 0.230942(1) & 0.5647(112) & -1.2457(199) &  5  &  $f_{3}-f_{1}$ \\
F8& 0.031280(1) & 0.5630(112) & -1.4853(200) &  4  &  $f_\mathrm{orb}$ \\
\hline   
\end{tabular} 

$^*$ denotes the frequencies used for the simultaneous binary light-curve, pulsation curve fitting process (see Sect.\,\ref{lcanalysis}).
\end{center}  
\end{table*}

\subsection{Discussion: Oscillations and Tidal Effects}
\label{Subsect:tidaloscillation}

The presence of oscillations at integer harmonics of the orbital frequency is not surprising. As described above, these oscillations are produced by a combination of tidal ellipsoidal effects, reflection effects, and Doppler beaming \citep[see][]{shporeretal11}. In a circular orbit, the reflection and Doppler beaming effects contribute mainly to variation at $f_{\rm orb}$, while the tidal effect contributes mainly to variation at $2 f_{\rm orb}$. For an eccentric orbit, each of these effects contributes variations at every harmonic of the orbital frequency \citep[see][]{wel11, burkartetal12}. In HD\,183648, the low eccentricity implies that these effects only contribute at low harmonics of the orbital frequency. Indeed, we only observe modulation at $f_{\rm orb}$, $2f_{\rm orb}$, and $3f_{\rm orb}$. In what follows, we assume all modulation arises from the primary since its light dominates the luminosity of the system.

The especially odd feature of HD\,183648 is the phase of the oscillation at $2f_{\rm orb}$. In typical nearly circular eclipsing binaries which show ellipsoidal variations, the eclipses occur near the minima of the ellipsoidal modulations, while the maxima occur one quarter of an orbital period later. The phase of this modulation is intuitive: the maxima occur away from eclipse when the equilibrium tidal distortion\footnote{The equilibrium tide is the hydrostatic tidal bulge raised on the star by the gravitational force of the companion star. The equilibrium tide creates a tidal bulge along the line connecting the center of mass of the two stars. Typically, the tidal bulge is decomposed into spherical harmonics. Here we consider only the dominant components of the equilibrium tide, namely the $l=2$, $|m|=2,0$ components.} causes the star to present a larger surface area toward the line of sight. However, in HD\,183648, the oscillation at $2f_{\rm orb}$ shows the opposite phase, with maxima near the eclipses (phases 0 and 0.5).

There are three possible explanations for the strange features of the oscillations at orbital harmonics in HD\,183648. The first is that non-adiabatic effects near the surface of the star are strongly affecting the temperature perturbation created by the equilibrium tidal distortion of the primary star. Tidal ellipsoidal variations are typically modelled by using Von Zeipel's theorem to calculate the surface temperature perturbations. In this case, the tidally depressed regions (where the surface gravity is stronger) are hotter, creating a luminosity fluctuation of the same phase as described above (i.e., the luminosity maxima occur away from eclipses). However, as shown in \citet{pfahletal08}, non-adiabatic effects can completely alter the temperature perturbations in hot stars with radiative envelopes like the primary in HD\,183648. For hot stars, the luminosity variation is typically dominated by temperature variations (rather than surface area distortion), and the phase of this variation can be arbitrary. Therefore, non-adiabatic effects may be strongly altering the luminosity variations produced by the equilibrium tidal distortion of HD\,183648, leading to the strange phase of the oscillation at $2 f_{\rm orb}$. 

A second explanation is that dynamical tidal effects are important. In stars with radiative envelopes, dynamical tides are composed of stellar g\,modes that are nearly resonant with the tidal forcing frequencies. As with the equilibrium tide, the dynamical tide produces observable oscillations at exact integer harmonics of the orbital frequency \citep{kumaretal95, wel11, fullerlai12, burkartetal12}. The phase of the luminosity fluctuations produced by dynamical tides can be different from that of the equilibrium tidal distortion \citep[see][]{olearyburkart14}, potentially creating the observed oscillations. However, in most cases the luminosity oscillations produced by the dynamical tide are smaller than that of the equilibrium tide \citep[see][]{tho12}, so a g\,mode unusually close to resonance may be needed to produce the oscillation at $2 f_{\rm orb}$. 

A third possibility is that non-linear interactions are affecting the mode phases and amplitudes. We discuss this in greater detail below.

\begin{figure}
\begin{centering}
\includegraphics[scale=.3]{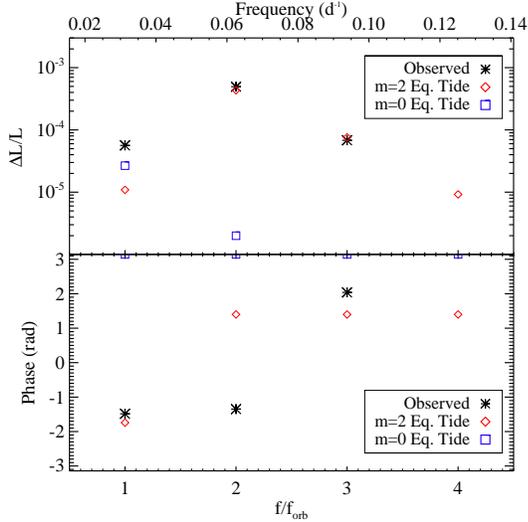}
\caption{\label{kic85plot} Top: Observed luminosity fluctuations and luminosity fluctuations due to the equilibrium tidal distortion (calculated using Von Zeipel's theorem) as a function of the orbital harmonic $f/f_{\rm orb}$. Bottom: Observed cosine phase of the luminosity fluctuations and the expected phase for the equilibrium tide. Although the observed oscillation amplitude at $2 f_\mathrm{orb}$ is near the expected equilibrium tide amplitude, it is  out of phase from the expected equilibrium tide phase.}
\end{centering}
\end{figure}

A full calculation of tidal excitation of non-adiabatic oscillation modes in rotating stars is beyond the scope of this paper. Instead, we simply calculate the expected luminosity fluctuation and phase of the adiabatic equilibrium tidal distortion, using the stellar parameters of Table\,\ref{Tab:syntheticfit}. The tidal distortion causes luminosity fluctuations of form
\begin{equation}
\label{eqtide}
\frac{\Delta L}{L} = A_n \cos \bigg(2 \pi n f_{\rm orb} t+ \delta_n \bigg),
\end{equation}
where the integer $n$ is the orbital harmonic of the oscillation, $A_n$ is its amplitude, and $\delta_N$ is its phase relative to periastron when $t=0$. We calculate the amplitude of the equilibrium tide as described in \citet{burkartetal12}, using Von Zeipel's theorem to calculate the flux perturbation. We also calculate the expected phase of the equilibrium tide luminosity fluctuation, which is\footnote{The phase can change by $\pi$ depending on whether the amplitude $A_n$ is positive or negative, which in turn depends upon, e.g., the sign of the Hansen coefficients used to calculate the tidal potential for eccentric orbits \citep[see][]{burkartetal12}. We calculate these coefficients, and adjust the phase $\delta_{n, \rm eq}$ such that the amplitude $A_n$ is positive.}
\begin{align}
\label{phase}
\delta_{n, \rm eq} &= |m| \omega \ \ {\rm for} \ |m|=2 \nonumber \\
&= \pi \ \ {\rm for} \ \ {\rm for} \ m=0 \, ,
\end{align}
where $\omega$ is the argument of periastron listed in Table\,\ref{Tab:syntheticfit}. We plot these results in Fig.\,\ref{kic85plot}. It is evident that although the magnitude of the observed luminosity fluctuations are similar to those expected from an adiabatic equilibrium tide, the phases are completely different. Hopefully, a more in depth investigation of tidally excited oscillations in this system will provide constraints on the tidal dynamics at play. 

Finally, we emphasize that it is important to consider the rapid rotation frequency of the primary in HD\,183648 relative to the orbital frequency. If the primary's spin axis is aligned with the orbit, the stellar rotation period is about 1.6\,d, implying the rotation frequency is $f_{\rm spin} \simeq 19.98 f_{\rm orb}$. In this scenario, the observed g\,modes ($f_1$ and $f_2$) cannot be prograde modes (in the rotating frame of the star) because their minimum frequency in the inertial frame is $|m| f_{\rm spin} > f_1, f_2$. Moreover, the tidally excited oscillations are retrograde oscillations in the rotating frame of the primary (although they are prograde in the inertial frame). Note that in the rotating frame, the tidal forcing frequencies are $f_{\rm tide} = n f_{\rm orb} - |m| f_{\rm spin}$. Since $f_{\rm spin} \gg f_{\rm orb}$ the absolute values of the forcing frequencies are much larger in the rotating frame, allowing for excitation of g\,modes with frequencies (in the rotating frame) $f_\alpha \sim |m| f_{\rm spin}$. 

\subsubsection{Non-linear Mode Coupling}

As described above, the dominant two oscillation frequencies $f_1$ and $f_2$ are separated by exactly $2 f_{\rm orb}$, which is the third largest amplitude oscillation observed. It is well known that combination frequencies of this sort are indicative of non-linear mode coupling \citep[see, e.g.,][]{wugoldreich01}. Indeed, there are now many cases of combination frequencies in close binaries which appear to be caused by non-linear mode coupling with tidally excited modes \citep[see][]{mukadametal10, fullerlai12, burkartetal12, ham13}.
In the case of HD\,183648, the non-linear coupling causes interactions between two g\,modes (corresponding to $f_1$ and $f_2$ in Table\,\ref{freqs}) and a tidally excited oscillation at $2 f_{\rm orb}$. At least one of the g\,modes may be self-excited, perhaps because one of the stars is a $\gamma$\,Dor variable. Indeed, the primary of HD\,183648 lies near the hot end of the $\gamma$\,Dor instability strip, while the secondary lies near the cool end. The observed frequencies $f_1$ and $f_2$ are on the low side, but are compatible with $\gamma$\,Dor pulsations \citep{balonaetal11}. The primary also lies within the $\delta$-Scuti instability strip, although no p-modes are observed.

The tidally excited oscillation is either composed of the equilibrium tide (which is dominated by the $l=2$, $|m|= 2$ f\,modes) or the dynamical tide (which is dominated by an $|m|=2$ g\,mode). The modes interact via a parametric resonance, redistributing energy amongst the three modes\footnote{In the classic parametric resonance discussed by, e.g., \citet{wugoldreich01}, a self-excited ``parent'' mode non-linearly transfers energy to two ``daughter'' modes. However, it is also possible for two self-excited (or tidally excited) parent modes to transfer energy to a single daughter mode.}. This transfer of energy changes the phases of the observed oscillations, and it may be possible that this is affecting the phase of the $2 f_{\rm orb}$ oscillation. 

It is also possible that $f_1$ and $f_2$ are not self-excited modes, but instead are non-linearly tidally driven modes. The signature of non-linear tidal excitation is that $f_{\alpha} \pm f_{\beta} = n f_{\rm orb}$ where $n$ is an integer \citep[see][]{weinbergetal12}. In HD\,183648, $f_2 - f_1 = 2 f_{\rm orb}$, which is compatible with non-linear excitation of $f_1$ and $f_2$ (although this does not explain the large amplitude of $f_3$). Non-linear tidal driving was observed in a similar system examined by \citet{ham13}.

\section{Summary and conclusions}

We have presented a complex photometric and spectroscopic analysis of HD\,183648, a marginally eccentric ($e = 0.05$), wide ($P_\mathrm{orb} = 31.973$), detached eclipsing binary system with a low amplitude pulsating component. The photometric analysis of the extremely accurate {\it Kepler} Q0 -- Q16 long cadence photometry incorporated disentangling of the eclipse and pulsation features, an eclipsing light curve solution and extended orbital period study via an ETV analysis. The spectroscopic investigations were based on ground-based high- and medium resolution spectra obtained with various instruments (echelle spectrograph at KPNO, ARCES Echelle spectrograph at APO, Hamilton Echelle Spectrograph at Lick Observatory, and eShel spectrograph of GAO, mounted on two telescopes at Szombathely and Piszk\'estet\H o, in Hungary) between 2011 and 2013. The spectroscopic data were mainly used for radial velocity analysis and for determination of stellar atmospheric properties and evolutionary states. Furthermore, the spectral disentangling technique made it also possible to detect the spectral lines of the secondary star despite its small (less than 5\%) contribution to the total light of the system. This fact allowed us to calculate dynamical masses and hence, relatively accurate stellar parameters. However, we emphasize that all of the various investigations were carried out in a complex and interdependent manner. Namely, the results and constraints of the radial velocity and ETV analysis were incorporated in the light curve analysis and vice versa, in an iterative manner; similarly, the quantitative spectral analysis was constrained at the same time by the outputs of the light curve analysis. In this way we were able to find a solution consistent with both the observations and the theoretical constraints. We found that the binary is composed of two main sequence stars with an age of $0.9\pm0.2$\,Gyr, having fundamental parameters of $M_1=1.93 \pm 0.126$\,M$_{\odot}$, $R_1=3.30 \pm 0.07$\,R$_{\odot}$ for the primary, and $M_2=1.06\pm 0.08$\,M$_{\odot}$, $R_2=1.11 \pm 0.03$\,R$_{\odot}$ for the secondary. Both stars were found to be rapid rotators with $\left(v_\mathrm{rot}\sin i_\mathrm{rot}\right)_1=104$\,km\,s$^{-1}$ and $\left(v_\mathrm{rot}\sin i_\mathrm{rot}\right)_2=26$\,km\,s$^{-1}$ which in the aligned case correspond to rotation periods $P_\mathrm{rot1}=1.60 \pm 0.04\,{\rm d}\sim19.98f_\mathrm{orb}$ and $P_\mathrm{rot2}=2.15 \pm 0.21\,{\rm d}\sim14.87f_\mathrm{orb}$, respectively.

We have found various types of eclipse timing variations in our analysis. We showed that the short time scale ($\sim\!287$\,d) periodic variation is a false positive due to an apparent beating between orbital and pulsational frequencies. The parabolic variation, which indicates a period increase with a constant rate, however, is suggested to be a real effect. The most plausible explanation is presence of an additional, distant, third body in the system. 

Clear indicators of apsidal motion have been found as well. We found a significant discrepancy between the theoretically computed and observed apsidal advance rates. This can be explained either with the insufficiently short time coverage of the apsidal motion cycle, which has a period of the order of ten thousand years, or perturbations from a tertiary component. Another alternative explanation is precession induced by dynamical tides \citep{willemsclaret05}. 

We made efforts to separate the oscillatory features from the binary characteristics, but there are strong connections between binarity and the detected oscillations. First, the difference of the two most dominant oscillation frequencies is equal to the twice of the orbital frequency, which indicates a binary origin. Furthermore, the most enigmatic feature of the out-of-eclipse part of the light curve is a sinusoidal variation with similar frequency and amplitude expected for the ellipsoidal effect, but with a completely opposite phase. Finally, there is a low amplitude oscillation at three times the orbital frequency, in addition to the \textquotedblleft inverse ellipsoidal\textquotedblright variation at twice the orbital frequency.

These phenomena are likely due to tidal effects. The oscillations at two and three times the orbital frequency are most likely tidally induced oscillations. However, it is unclear whether they are produced by hydrostatic equilibrium tides or by tidally excited g-modes. If they are equilibrium tides, non-adiabatic effects must be strongly altering their observed phase. If they are g-modes, they must be resonantly excited to account for their large amplitudes. Finally, the observed combination frequency $F_2 - F_1 = F_3 = 2 f_{\rm orb}$ indicates that non-linear mode coupling with the tidally excited oscillations is occurring. We are hopeful that a more detailed tidal analysis of HD 183648 may explain these observations and yield constraints on tidal dissipation theories. 

\section*{Acknowledgements}

This project has been supported by the Hungarian OTKA Grant K83790, ESA PECS Contract No. 4000110889/14/NL/NDe, the Lend\"ulet-2009 Young Researchers Programme of the Hungarian Academy of Sciences and the European Community's Seventh Framework Programme (FP7/2007-2013) under grant agreement no. 269194 (IRSES/ASK) and no. 312844 (SPACEINN). AD, RSz and GyMSz have been supported by the J\'anos Bolyai Research Scholarship of the Hungarian Academy of Sciences. TB, BCs, JK and GyMSz would like to thank City of Szombathely for support under Agreement No. S-11-1027. Based on observations obtained with the Apache Point Observatory 3.5-meter telescope, which is owned and operated by the Astrophysical Research Consortium.

\end{document}